\documentclass[ aip, jmp, amsmath,amssymb,reprint,]{revtex4-1}

\usepackage{graphicx}
\usepackage{dcolumn}
\usepackage{bm}

\newcommand{\be}{\begin{equation}}
\newcommand{\ee}{\end{equation}}
\newcommand{\beq}{\begin{equation}}
\newcommand{\eeq}{\end{equation}}
\newcommand{\bea}{\begin{eqnarray}}
\newcommand{\eea}{\end{eqnarray}}
\newcommand{\nn}{\nonumber}

\newcommand{\bear}{\begin{eqnarray}}
\newcommand{\eear}{\end{eqnarray}}

\def\cb{{\cal B}}
\def\cc{{\cal C}}
\def\cph{{\mathbf{ \varphi}}}
\def\ch{{\cal H}}

\def\cz{{\cal Z}}

\def\pperp{{\perp\perp}}
\def\to{\rightarrow}
\newcommand{\qn}{\mathfrak{q}}
\newcommand{\wn}{\mathfrak{w}}

\begin{document}

\preprint{ITP-UU-11/35, DFTT 27/2011}

\title[~~~~]{Holographic Duals of Quark Gluon Plasmas with Unquenched Flavors}

\author{Francesco Bigazzi}
 \email{bigazzi@fi.infn.it}
\affiliation{Dipartimento di Fisica e Astronomia, Universit\`a di Firenze; Via G. Sansone 1, I-50019 Sesto Fiorentino (Firenze), Italy, and INFN, Sezione di Pisa, Largo B. Pontecorvo, 3, 56127 Pisa, Italy.
}
\author{Aldo  Cotrone}%
 \email{cotrone@to.infn.it}
\affiliation{
Dipartimento di Fisica, Universit\'a di Torino and INFN Sezione di Torino; \\Via P. Giuria 1, I-10125 Torino, Italy.
}%

\author{Javier Mas}
 \email{javier.mas@usc.es}
\affiliation{%
 Departamento de  F\'\i sica de Part\'\i culas, Universidade de Santiago de Compostela and Instituto Galego de
F\'\i sica de Altas Enerx\'\i as (IGFAE); E-15782, Santiago de Compostela, Spain.
}%

\author{Daniel Mayerson}
 \email{d.r.mayerson@uva.nl}
\affiliation{%
Institute for Theoretical Physics, University of Amsterdam,
Science Park 904, Postbus 94485, 1090 GL Amsterdam, The Netherlands. 
}%

\author{Javier Tarrio}
 \email{l.j.tarriobarreiro@uu.nl}
\affiliation{%
Institute for Theoretical Physics, Universiteit Utrecht, 3584 CE, Utrecht, The Netherlands.
}%

\date{\today}

\begin{abstract}
We review the construction of gravitational solutions holographically dual to $\mathcal{N}=1$ quiver gauge theories with dynamical flavor multiplets. We focus on the D3-D7 construction and consider the finite temperature, finite quark chemical potential case where there is a charged black hole in the dual solution. Discussed physical outputs of the model include its thermodynamics (with susceptibilities) and general hydrodynamic properties.
\end{abstract}

\maketitle

\tableofcontents

\section{Introduction}

Despite the fact that the QCD Lagrangian can be written down in a line, deriving the huge plethora of strongly interacting phenomena that are seen in accelerators (and in everyday life) has remained a major challenge for over half a century.
Especially at low energies, QCD still remains intractable by traditional perturbative methods.
A new window into QCD physics has recently opened after the experiments carried out at RHIC during the last decade.\cite{arsene} Now, the new collision experiments at RHIC and LHC (see for example Ref. \onlinecite{lhc} for early results) will allow obtaining much more precise informations about the phase diagram of QCD
in the region of high temperature and small baryon density. During the last years, evidence of the formation of a quark-gluon plasma has been growing, in the form of a transient object called the ``fireball'',  which expands and eventually explodes into hadrons.\cite{arsene}\\

 String theory, which was originally devised   to explain the low energy phenomenology of QCD, has ended up producing one of the most valuable theoretical methods to explore strongly coupled systems: the $AdS$/CFT duality. While its prototypical example, ${\cal N}=4$ supersymmetric Yang-Mills theory, is far from being a realistic analog of QCD, its finite temperature counterpart shares with QCD many properties such as Debye screening, a deconfined phase, etc. It was an impressive unexpected result to find that  the ratio of viscosity to entropy density in the ${\cal N}=4$ SYM plasma at strong coupling turned out to be an almost universal prediction, whose value comes closest to the results that fit the data obtained from RHIC and LHC.\\

 The aforementioned breakthrough boosted an enormous activity, and opened up a very interesting field of research that attracted both physicists from string theory as well as from heavy ion and QCD communities. The core of this program is to obtain the physical transport properties of a relativistic quark gluon plasma
(much of the research in this field is discussed at length in the recent review in Ref. \onlinecite{CasalderreySolana:2011us}, which also contains an extensive reference list).
In its more crude form, the initial attempts were really only dealing with adjoint degrees of freedom. To say that $N_c=\infty$ is a sensible approximation to $N_c=3$ is probably not as severe a statement as to say
 that in real QCD the number of flavors is zero, or even negligible as compared with the number of colors.\\

From the holographic perspective, adding flavor degrees of freedom that transform in the fundamental representation of the gauge group
is  fairly well understood  in the so called {\em quenched} approximation. A  quenched quark,  a term borrowed from the lattice literature,  means a non-dynamical quark. In perturbation theory  diagrams with internal quark loops in it are neglected. The easiest realization of this is the case of very massive quarks.
For light quarks there is another {\em quenching} mechanism in the so called 't Hooft limit,\cite{'tHooft:1973jz} $N_c\to \infty$ and $\lambda = g_{YM}^2 N_c$ finite. In such a limit, diagrams with internal quark loops are suppressed by factors of $N_f/N_c$.\cite{Veneziano:1976wm,Capella:1992yb} Therefore keeping $N_f$ fixed induces a {\em quenching} effect even for massless quarks.\\

From the gravity dual perspective, this approach amounts to having a finite number of \emph{probe} flavor branes which do not backreact onto the geometry that the color branes create. There have been several attempts to go beyond the quenched approximation by considering metrics sourced by an infinite number of $N_c$ color and $N_f$ flavor branes. Apart from the works explicitly discussed in the rest of this paper, we refer the reader to Refs. \onlinecite{Casero:2005se,tutti} and to the excellent review in Ref. \onlinecite{review} for an extensive list of references on the topic. The present review will be concerned exclusively with one such strategies, which we will term as ``\emph{smeared flavors}". Other proposals where the backreaction of a localized stack of D7-branes is taken into account  are included in Ref.  \onlinecite{localizedD3D7}. \\

{\bf Outline:}
 Our aim in this review is to provide a very general account of the so called D3-D7 holographic quark gluon plasmas, including
 finite temperature and chemical potential dependence.\cite{D3D7QGP,hydro1,hydro2,Bigazzi:2011it}
 In section II we review the essential ingredients of the ``smeared flavor" construction, in the particular setup of D7-branes in the background of D3-branes. The geometry sourced by the D3-branes will be of the general type having a Sasaki-Einstein internal space. In section III we introduce the ansatz that incorporates the desired physics, and the equations of motion are derived.  Section IV is devoted to solutions to these equations. An analytic solution is possible as a perturbative expansion in small values of $\epsilon \sim N_f/N_c$ and the baryon density $n_b$. In essence we extend and fill in some gaps of the work presented in Ref. \onlinecite{Bigazzi:2011it} and provide the explicit solution up to  second order  in  $\epsilon$ and first nontrivial order in $n_b$. In section V we shall show how to obtain consistent thermodynamics from this solution and discuss the susceptibilities. Finally, in section VI we explore some other physical predictions of the mentioned model. In particular, the hydrodynamic and optical properties of the quark gluon plasmas are examined.
We conclude with some appendices containing complementary material.

\section{D3-D7-brane systems}
Introducing a stack of $N_c$ D3-branes at the tip of a 6d Calabi-Yau cone over a 5d Sasaki-Einstein manifold $X_5$ provides a gravity theory dual to a supersymmetric field theory in 4d which preserves (at least) $\mathcal{N}=1$ supersymmetry (see Refs. \onlinecite{Maldacena:1997re,Klebanov:1998hh}; for a review see Ref. \onlinecite{MAGOO}). By far the most studied example of this setting is when the Sasaki-Einstein manifold is $S^5$, so that the gravity theory is that of $AdS_5\times S^5$, and the dual field theory is the conformal field theory of $\mathcal{N}=4$ SYM with $SU(N_c)$ gauge symmetry. However, these theories do not yet include  matter in the fundamental representation of the gauge group; to do this, we must deform the theory by adding the flavor degrees of freedom. This section reviews briefly the probe approximation and the smearing approximation, two ways in which flavor can be added to the field theory.

\subsection{Introducing Flavors}
We can add  matter in the fundamental representation of the $SU(N_c)$ gauge group  by introducing D7-branes into the system; a stack of $N_f$ coincident D7-branes  introduces a $U(N_f)$ global flavor symmetry into the theory. Then, we have three types of open string excitations: open strings ending on two D3-branes which describe gluons in the adjoint of $SU(N_c)$, open strings ending on both a D3 and a D7-brane which describe quarks in the fundamental of $SU(N_c)$ and open strings ending on two D7-branes. The latter are in the adjoint representation of $U(N_f)$ and thus naturally represent mesons (for a review, see Ref. \onlinecite{Erdmenger:2007cm}).\\

Adding D7-branes in this fashion was first done for $\mathcal{N}=4$ $SU(N_c)$ SYM in Refs. \onlinecite{Kehagias:1998gn}, \onlinecite{Aharony:1998xz}, \onlinecite{Karch:2002sh}. The introduction of flavors in this way breaks the $\mathcal{N}=4$ supersymmetry to $\mathcal{N}=2$ and also  conformality in the field theory. The following intersection of the D3 and D7-branes is used
{
\begin{center}
\begin{tabular}{|c|c|c|c|c|c|c|c|c|c|c|}
\hline
 & $t$ & $x^1$ & $x^2$ & $x^3$ & $r$ & $a^1$ & $a^2$  & $a^3$ & $a^4$ & $a^5$\\
\hline
 D3 & X & X & X & X & & & & & & \\
\hline
 D7 & X & X & X & X & X & X & X & X &  & \\
\hline
 \end{tabular}
 \end{center}
}
Here, $a^{1-5}$ are the coordinates on the internal compact manifold $X_5$. The D3-branes are localized along the $AdS$ radial variable $r$ at $r=0$. If we want massless quarks, the D7-branes should extend to the origin and intersect the D3-branes; for massive quarks, the D7-branes should have a radial profile and only extend down to a certain radial position $r_q$ - this position is then related to the mass of the quarks.
We will be mainly concerned with massless quarks in the following.\\

The contribution to the full action of the flavor D7-branes compared to the contribution of the D3-branes can be calculated as\cite{review}
\be
\mathcal{L}_{D7} \sim \frac{N_f}{N_c} \lambda\  \mathcal{L}_{D3}\, ,
\ee
where $\mathcal{L}_{D7,D3}$ are the respective Lagrangians and we have used the 't Hooft coupling $\lambda= g_{YM}^2 N_c$. Essentially, this implies that the parameter $N_f/N_c$ (or, to be more precise, $\lambda N_f/N_c$) is the measure of the backreaction of the D7-branes on the $AdS$ background geometry sourced by the D3-branes. In fact, the backreaction of the $N_f$ D7-branes is zero in the 't Hooft scaling limit (keeping $N_f$ fixed):\cite{'tHooft:1973jz, review, Erdmenger:2007cm}
\be \label{D3D7:thooft}
\frac{N_f}{N_c}\rightarrow 0\, ,
\ee
i.e. when we take the number of flavors to be very small compared to the number of colors. This is called the \emph{probe} approximation for the D7-branes, as they probe the geometry sourced by the D3-branes without deforming it. From the field theory perspective, we are ignoring the quantum effects caused by the quarks - this can be seen by considering different diagrams contributing to a given physical process: the contribution of a diagram with $k$ internal quark loops compared to a similar diagram with none scales as $(N_f/N_c)^k$. Thus, using the probe approximation for the D7-branes gives us quarks in the dual field theory which are essentially external non-dynamical objects which do not run in loops. This is the \emph{quenched} approximation for the fundamental matter in the field theory, a term borrowed from lattice literature; the approximation becomes exact in the limit (\ref{D3D7:thooft}).\\

The dynamics of the D7-brane in the probe approximation is governed by its DBI action,
\be
S_{D7} = -T_7 \int d^8 \xi e^{-\Phi} \sqrt{\det( \hat G_{ab} + \hat B_{ab} + 2\pi\alpha' F_{ab})}\, ,
\ee
where $\hat G_{ab}$ is just the pullback of the $AdS_5\times X_5$ metric sourced by the D3-branes. In the probe approximation, one only needs to consider this action to determine the full dynamics of the D7-branes.

\subsection{Finite Baryons Density}
QCD is expected to show a rich phase diagram at finite baryon density, including a color-flavor locking phase at high baryon density and relatively low temperature. In addition, neutron stars are astronomical objects expected to be at virtually zero temperature but non-zero baryon density. Thus, it would be interesting to use holography to probe properties of strongly interacting field theories like QCD at finite baryon density.\\

In QCD, the subgroup $U(1)_B$ of the $U(N_f=3)$ global flavor symmetry is identified with the baryon number. Thus, we wish to introduce a finite $U(1)_B$ density $\langle J^t \rangle$, where $J^{\mu}$ is the $U(1)_B$ current. This baryonic current is dual to a vector field $A_{\mu}$ living on the D7-branes.\cite{Kim:2006gp,hep-th/0611099, Karch:2007br} The precise correspondence is, as calculated in the probe analysis:\cite{hep-th/0611099}
\bea \nn  A_t(r) \leftrightarrow \mathcal{O}_q & = & \psi^{\dagger}\psi + \tilde{\psi}\tilde{\psi}^{\dagger}+ i\left(q^{\dagger}\mathcal{D}_t q-(\mathcal{D}_t q)^{\dagger} q\right)\\
\label{D3D7:Atholodictionary} && + i\left(\tilde{q}(\mathcal{D}_t \tilde{q})^{\dagger}-\mathcal{D}_t \tilde{q}\, \tilde{q}^{\dagger}\right) ,\eea
where $\psi,\tilde{\psi}$ are the fermions in the (anti)fundamental representation of $SU(N_c)$, and $q,\tilde{q}$ are scalars in the (anti)fundamental representation; $\mathcal{D}$ is the covariant derivative in the correct representation; $\mathcal{O}_q$ is the quark number operator. Note that we only need the time component of the gauge field $A_t$, and that this field only depends on the $AdS$ radius $r$.
A complete analysis of the stability of the system in the probe approximation can be found in Ref. \onlinecite{Ammon:2011hz}, while recent studies for imaginary chemical potential are reported in Ref. \onlinecite{Aarts:2010ky}.\\

In the probe analysis, the asymptotic behavior of $A_t(r)$ for large $r$ is\cite{hep-th/0611099}
\be
A_t(r) \sim \mu - \frac{a}{r^2}\, ,
\ee
where $\mu$ is the quark chemical potential and $a$ is proportional to the quark density $n_q$. The quark density $n_q$ is related to the baryon density $n_B$ as $n_B = n_q/N_c$ as there are $N_c$ quarks in a baryon; analogously $\mu_B = \mu N_c$.\\

In Ref. \onlinecite{matsuuracfl}, the so-called linearized backreaction on the supergravity geometry of including a non-zero $A_t(r)$ is calculated. The result is that a component of the RR-field $F_{(3)}$ is turned on (see also Ref. \onlinecite{Callan:1986bc}). For the zero temperature case, it is found that $F_{(3)} = F_{123} dx^1\wedge dx^2\wedge dx^3$, where
\be
F_{123} = \frac{8\pi^3 \alpha'^2 d}{N_c}\, .
\ee
Here $d$ is a constant related to the quark density as $d=n_q/\mathrm{Vol}(S^3)$ and is obtained as a constant of motion from $\delta {\cal L}_{D7}/\delta F_{rt}$, with $F_{rt}=\partial_r A_t$.\cite{hep-th/0611099, matsuuracfl} This constant $F_{123}$ component of $F_{(3)}$ will be an important motivation for the choice of ansatz for $F_{(3)}$ below.


\subsection{Beyond Quenched Flavors}
The discussion in this section is mainly inspired by that in Refs. \onlinecite{review} and \onlinecite{Benini:2006hh}. We shall keep the discussion fairly minimal here, we refer the reader to Ref. \onlinecite{review} for an in-depth review of the smearing procedure for backreacting D7-branes.\\

As discussed above, in the 't Hooft scaling limit (\ref{D3D7:thooft}), the backreaction of the D7-branes can be neglected. However, if we want to introduce dynamical flavors that run in loops in our field theory, this scaling limit is insufficient; we need to take the backreaction of the flavor branes into account. We can do this by using the Veneziano limit, \cite{Veneziano:1976wm, review}
\be
N_f\rightarrow \infty, \qquad \frac{N_f}{N_c} = const.
\ee
Thus, now we keep the ratio $N_f/N_c$ fixed while taking both $N_f$ and $N_c$ to infinity. The ratio $N_f/N_c$ essentially determines the magnitude of the backreaction of the flavor branes, as we will see.\\

Clearly, we can not neglect the backreaction of the $N_f$ branes on the $AdS$ geometry in this scaling limit. Now, instead of considering the $AdS$ background as a solution to the IIB supergravity field equations and then minimizing the D7-brane DBI action in this geometry, we will minimize the full supergravity action $S_{SUGRA}$ \emph{together} with the D7-brane DBI and WZ action $S_{D7}$
\be
S = S_{SUGRA} + S_{D7}\, .
\ee
The D7-brane action $S_{D7}$ is an integral over the 8d worldvolume of the D7-branes. Such integrals are typically very difficult to deal with; this is why the powerful technique of smearing was introduced, which we will now discuss.

\subsection{Smearing}
Let us consider the Abelian DBI action (without form fields),
\be
S_{DBI} = \int_{\mathcal{M}_8} d^{8} \xi \sqrt{-\hat G}\, ,
\ee
which is an integral over the 8d worldvolume of the D7-brane. To write this as an integral over the entire 10d space-time, we can use a localized two-form $\Omega_2$ such that
\be\label{omegadue}
\Omega_2 =  \delta^{(2)}(\mathcal{M}_8) \frac{\sqrt{-\hat G}}{\sqrt{-g}} \alpha\wedge \beta\, ,
\ee
where $\alpha, \beta$ are one-forms which are orthogonal to $\mathcal{M}_8$ and $g$ refers to the ten-dimensional metric. Then we have
\be \label{D3D7:smearDBI}
S_{DBI} = \int_{\mathcal{M}_8} d^{8} \xi \sqrt{-\hat G} = \int d^{10}x \sqrt{-g} |\Omega_2| \, .
\ee
Note that $|\Omega_2|=\sum_i |\Omega^{(i)}|$ (where $\Omega^{(i)}$ are the decomposable parts of $\Omega_2$) is the coordinate-invariant modulus of $\Omega_2$, which in the supersymmetric or trivial embedding case reads
\be
|\Omega^{(i)}| = \sqrt{\frac{1}{2!} \Omega_{MN}^{(i)} \Omega_{PQ}^{(i)} g^{MP} g^{NQ}}\, .
\ee
Since the $\delta$-functions in (\ref{omegadue}) are virtually impossible to work with in the general case, it would be advantageous to get rid of them; this is precisely the purpose of the smearing procedure. One can effectively view this as taking a large amount of D7-branes and smearing them over their transverse space.\cite{Bigazzi:2005md}\\

\begin{figure}[h]
\centering
\begin{minipage}[t]{0.145\textwidth}
\begin{center}
\includegraphics[width=0.85\textwidth]{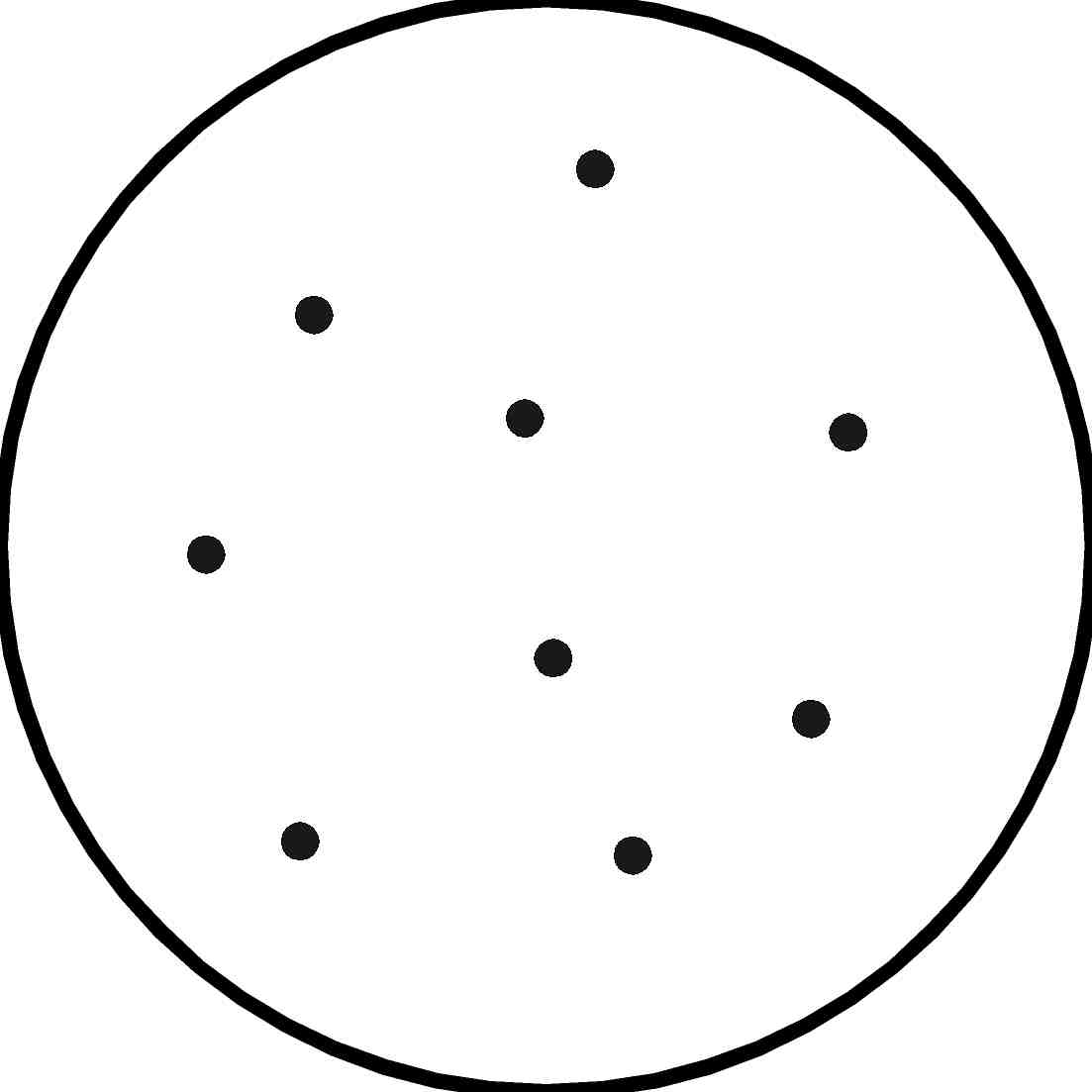}
\caption{A disc with a small number of points in it.}
\label{fig:D3D7discsmallnr}
\end{center}
\end{minipage}
\begin{minipage}[t]{0.01\textwidth}\hspace{0.5pt} \end{minipage}
\begin{minipage}[t]{0.145\textwidth}
\begin{center}
\includegraphics[width=0.85\textwidth]{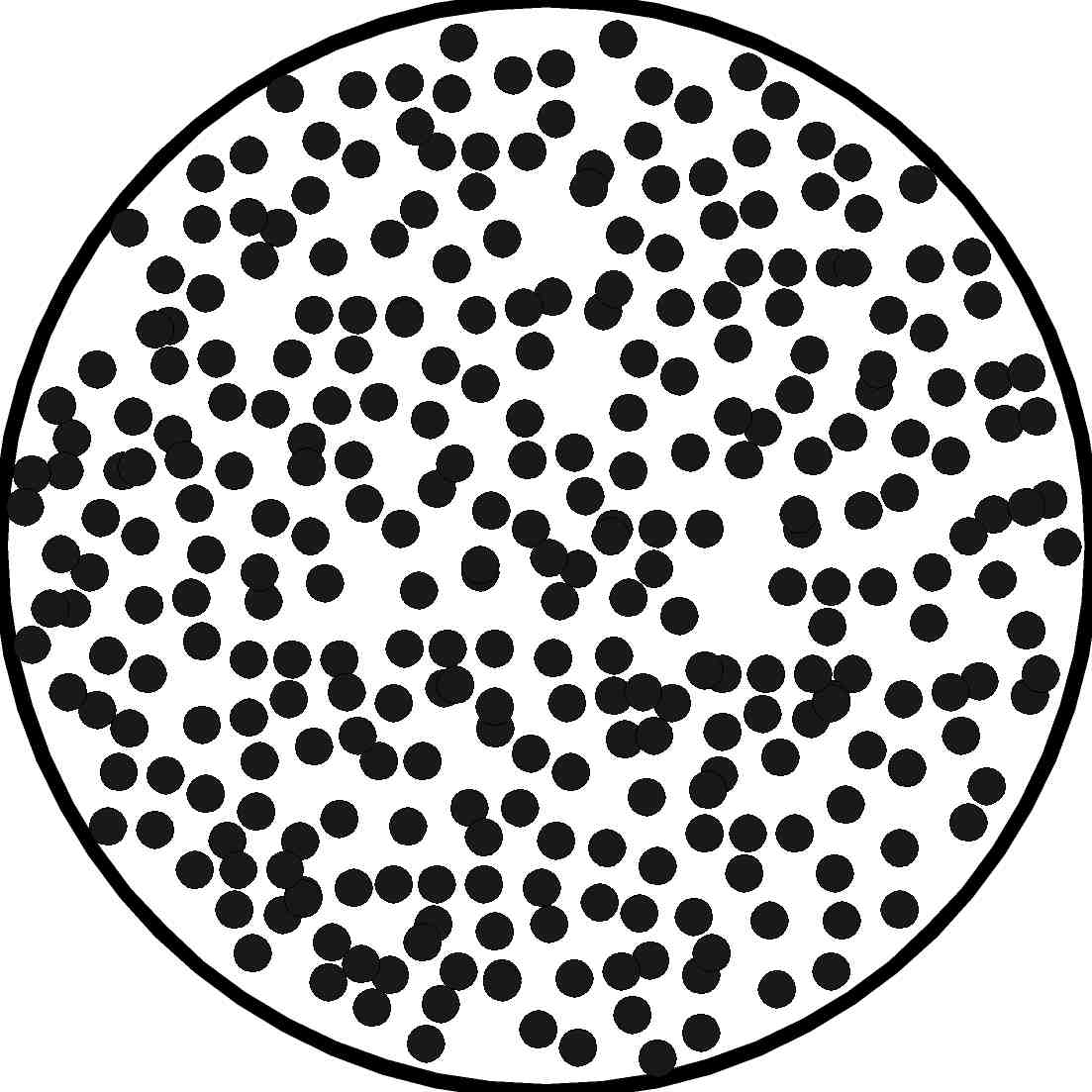}
\caption{A disc with a large number of points in it.}
\label{fig:D3D7disclargenr}
\end{center}
\end{minipage}
\begin{minipage}[t]{0.01\textwidth}\hspace{0.5pt} \end{minipage}
\begin{minipage}[t]{0.145\textwidth}
\begin{center}
\includegraphics[width=0.85\textwidth]{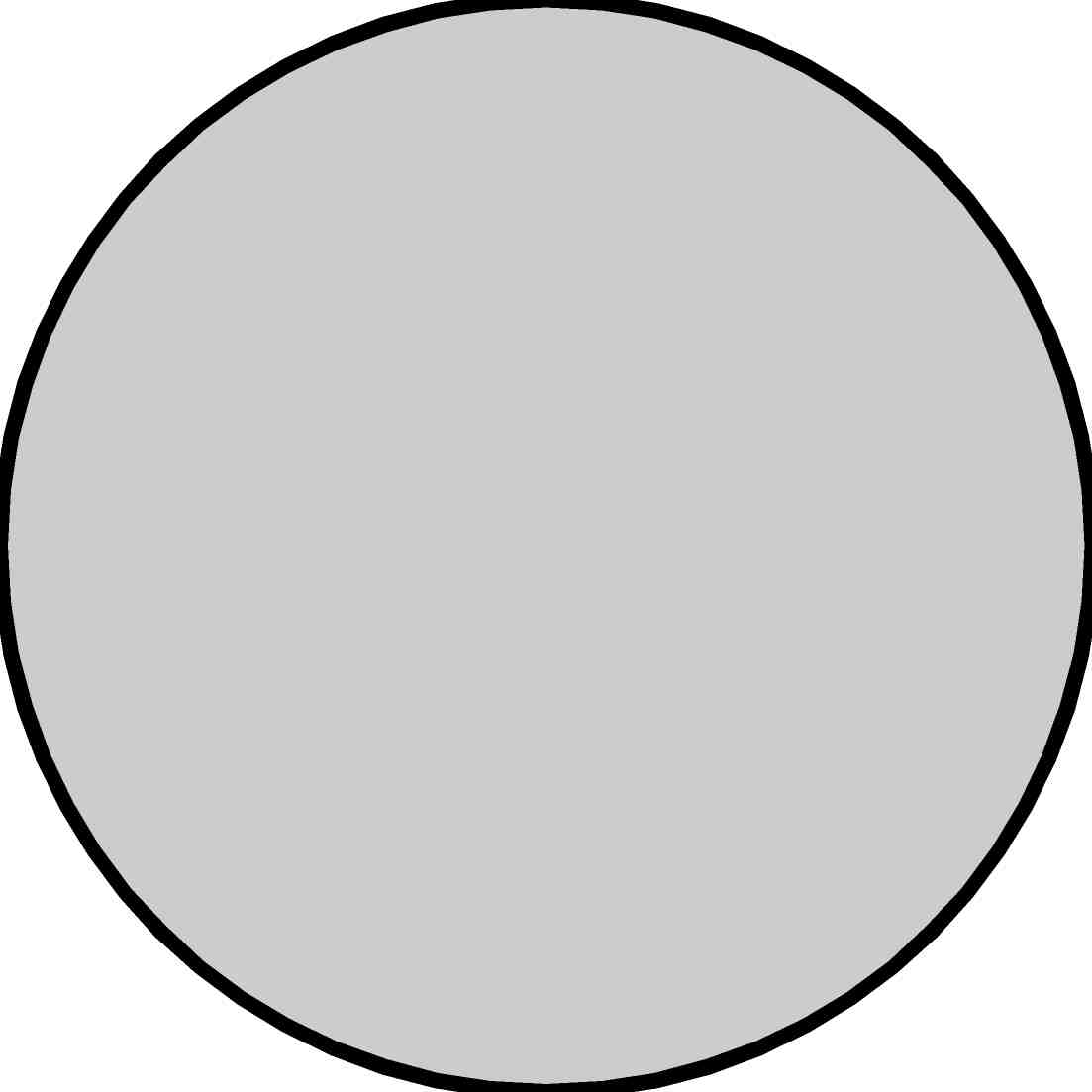}
\caption{The large number of points in the disc have been `smeared' out into a uniform distribution.}
\label{fig:D3D7discsmooth}
\end{center}
\end{minipage}
\end{figure}

An apt analogy for this smearing procedure using electrically charged lines is given in Ref. \onlinecite{review}. Let us here give our own short analogy. Consider a disc containing a number of points. If the number of points in the disc is small (see fig. \ref{fig:D3D7discsmallnr}), then we safely approximate the disc as having nothing in it - this is comparable to the probe approximation, where the number of D7-branes (the points) is small compared to the number of D3-branes (the disc's area). However, when the number of points is large (see fig. \ref{fig:D3D7disclargenr}), then we can no longer purport the disc to be ``almost'' empty - we have to take the points into account in the description of the disc. Taking each individual point into account would be quite a chore, indeed. However, if the points are relatively homogeneously spread throughout the disc, then we can approximate the distribution of points using a continuous distribution function - this turns our complicated disc in fig. \ref{fig:D3D7disclargenr} into the smooth, more manageable, disc in fig. \ref{fig:D3D7discsmooth}. We have now effectively ``smeared'' the points in the disc.
\\

It is in any case obvious that such smearing procedures are not by any means new in physics. For example, a liquid is really a collection of small particles, but from a macroscopic viewpoint we can `smear' the particles and just consider the fluid as a continuum. Note that for the fluid, it is important that there are a large number of particles; the situation is the same when we consider D7-brane smearing - we need a large number of them before the smearing procedure gives us a good approximation.\\

The two-form $\Omega_2$ as defined above essentially tells us where the D7-branes are located. Without smearing, it consists of delta functions and is not easy to work with; the smearing procedure tells us how to `smear' this two-form. The crucial observation in Refs. \onlinecite{cnp}, \onlinecite{Benini:2006hh} is now that a smeared $\Omega_2$ is not arbitrary if we want to preserve supersymmetry. It turns out we require, for massless flavors,
\be \label{D3D7omega2}
\Omega_2 = -2Q_f J_{KE}\, ,
\ee
where $J_{KE}$ is the K\"ahler two-form of the K\"ahler-Einstein base of the compact Sasaki-Einstein manifold $X_5$, and $Q_f$ is a constant related to $N_f$ as
\be Q_f = \frac{\mathrm{Vol}(X_3) g_s N_f}{4 \mathrm{Vol}(X_5)}\, ,
\label{flavdens}\ee
where $\mathrm{Vol}(X_5)$ is the volume of the internal manifold, and $\mathrm{Vol}(X_3)$ is the volume of the submanifold $X_3$ of $X_5$ that the D7-branes wrap. Note that every D7-brane in principle has a different manifold $X_3$ which it wraps, but all of these $X_3$'s are related to each other by symmetry transformations, so $\mathrm{Vol}(X_3)$ and thus $Q_f$ is well-defined.\\

Practically, we use the smeared $\Omega_2$ by making the following substitution for any eight-form $X_{(8)}$ which we integrate over the D7-brane worldvolume
\be \label{D3D7:smearingX8}
\int_{\mathcal{M}_8} X_{(8)} \rightarrow \int_{\mathcal{M}_{10}} X_{(8)}\wedge \Omega_2\, .
\ee
In particular, for pull-back quantities such as in the Wess-Zumino part of the D7-brane action, we have
\be
\int \hat{C}_{(8)} \rightarrow \int C_{(8)}\wedge \Omega_2\, .
\ee

We note that $\Omega_2$ can be seen as a magnetic source for the form-field $F_{(1)}$
\be
d F_{(1)} = -\Omega_2\, ,
\ee
since the D7-branes source an eight-form potential. This means that we can take
\be
F_{(1)} = Q_f (d\tau + A_{KE})\, ,
\ee
where $\tau$ is the $U(1)$ fibration coordinate of the Sasaki-Einstein manifold $X_5$, and $A_{KE}$ is the K\"ahler potential ($ dA_{KE} = 2 J_{KE}$).\cite{Benini:2006hh} We also note that we can give the quarks mass by giving the D7-branes a non-trivial profile in the radial direction; we can do this by introducing a function $p(r)$:
\be
F_{(1)} = Q_f p(r) (d\tau + A_{KE})\, .
\ee
Of course, $p(r)\rightarrow 1$ for $r\rightarrow\infty$, as the quarks are effectively massless in the UV at energies much higher than the quark masses; and $p(r)=0$ for $r<r_q$ for a certain radius $r_q$ as the quarks can be effectively integrated out at energies much lower than the quark masses. The function $p(r)$ can be calculated explicitly and can be shown to be a smoothed-out theta-function.\cite{Bigazzi:2008zt}

\subsection{Validity of the smearing}\label{sectionvalidity}

The smearing technique described above makes use of the Abelian  DBI action in order to take into account flavor effects.
This approach has some limitations; we discuss in this section three topics which sharpen its regime of validity.\\

As a first point,\cite{Bigazzi:2008zt,papadimitriou} the DBI action has been shown in Ref. \onlinecite{Callan:1986bc} to take into account the leading contribution of the open string sector, i.e. the ${\cal O}(g_sN_f)$ terms in the topological expansion of the string amplitudes.
This allows to resum, in the dual field theory, the leading effects in $N_f/N_c$, the ``one-window" graphs in the language of Ref. \onlinecite{Veneziano:1976wm}, i.e. the graphs containing one loop of matter in the fundamental representation (``quark loops").
The result is surely an improvement with respect to the probe limit, which does not include any loop of fundamental matter.
It does not provide the full backreaction of flavors, though.
The latter would require to resum the contributions from the graphs with any number of quark loops.
Going beyond the ``one-window" graph contributions is surely an interesting issue; a proposal can be found for example in Ref. \onlinecite{Gadde:2009dj}.
Nevertheless, if $g_sN_f$ is small, the smearing technique provides the leading flavor contributions, since the graphs with $k$ quark loops are weighted by powers of $(g_sN_f)^k$.
As we will see in the following, we will indeed require $g_sN_f$ to be small.
Note that in any case the effects taken into account by the DBI are quite non-trivial, non-linear in $g_sN_f$ because the ``one-window" graphs are resummed.\\

On top of this, the use of the {\emph{Abelian}} DBI is fully justified only if the minimal distance between two branes is much larger than the string length, i.e. the strings connecting different branes are massive and can be ignored among the light modes.
If $R$ is the typical size of the $n$ internal directions along which the smearing is performed, the minimal distance between two branes will be of order $R^n/N_f$.
This poses the restriction $N_f \ll R^n$.
In the example of this paper, $R^4 \sim \lambda, n=2$, and considering also the requirement $N_f \gg 1$, it turns out that the smearing approximation is justified in the parameter window $1 \ll N_f \ll \sqrt{\lambda}$.
Since in this window the non-diagonal modes among the D-brane spectrum are essentially decoupled, the smearing technique describes the flavor-singlet sector of the dual field theory.
An improvement in this direction could only be captured by a consistent non-Abelian DBI action.\\

Finally, let us point out a limitation concerning the use of the DBI in the finite temperature case.
It has been known for a while (for recent work in this direction see for example Ref. \onlinecite{Grignani:2010xm}) that the DBI does not capture the thermalization effects of the theory on the world-volume of the D-branes: its energy momentum tensor does not coincide with the black brane one.
As such, it constitutes only an approximation to the dual description of a fully thermalized gauge theory with fundamentals.
The effects that are not taken into account scale as $N_f/N_c$ times some power of the temperature (in the zero temperature limit the discrepancy is of course absent).
So, the DBI is a good approximation in the probe approximation or at small temperatures.
Thus, we will be describing the small temperature regime of the dual field theory; this limitation is anyway imposed upon us also by another reason: the presence of a Landau pole in the UV in the theories we will be considering.\\

To summarize, smearing the D7-brane action is a powerful technique which simplifies the D7-brane action to only depend on the $AdS$ radial coordinate; our equations of motion will (usually) thus be ODEs instead of PDEs, and in any case result in equations without $\delta$-functions. However, in addition to the validity limitations as described above, we must also beware that the smearing is only a good approximation if the number of flavor branes is large and moreover they are uniformly distributed throughout the directions transverse to the branes. So, the global $U(N_f)$ flavor symmetry sourced by $N_f$ coincident D7-branes is broken to $U(1)^{N_f}$ when we smear them, as smearing requires the D7-branes to sit at different points in the transverse space.

\section{Set Up \& Ansatz}
We wish to study a field theory which is dual to the Type IIB gravity theory on $AdS_5\times X_5$. This field theory must be deformed by $N_f$ dynamical flavors, so we will add a large number of D7-branes to the gravity theory, smeared in the way discussed above. Finally, we turn on a finite temperature and finite baryon density in this field theory; in the gravity theory this translates as the introduction of a black hole in the geometry and a gauge field $A_t$ on the D7-branes, respectively.\\

The method we must use to study this gravity theory is as in Refs. \onlinecite{D3D7QGP} and \onlinecite{Bigazzi:2011it}, i.e. the second order equations of motion must be solved for. We can not use supersymmetry considerations as we are interested in the system at finite temperature so that supersymmetry is broken.\\

The starting point is thus the full supergravity action together with the D7-brane action. This action is
\begin{widetext}
\bea
 \nn S & = & \frac{1}{2\kappa_{10}^2} \int d^{10}x \sqrt{-g} \left[ R - \frac12 \partial_M \Phi \partial^M \Phi - \frac12 e^{-\Phi} H_{(3)}^2 - \frac12 e^{2\Phi} F_{(1)}^2 -\frac12 e^{\Phi} F_{(3)}^3- \frac14 F^2_{(5)}\right]\\
\label{setup:theaction} && -\frac{1}{4\kappa_{10}^2}\int C_{(4)} \wedge H_{(3)} \wedge F_{(3)} + S_{fl}\, ,\\
\label{setup:flavoraction} S_{fl} & = &  - \mu_7 \int_{D7} d^8 \chi e^{\Phi} \sqrt{-\det (\hat{G} + e^{-\Phi/2} \mathcal{F})} + \mu_7\int \hat{C_q}e^{-\mathcal{F}} \, .
\eea
\end{widetext}
Here we have defined $\mathcal{F}:=2\pi\alpha' F + \hat{B}$. Note that $F_{(p)}^2 = \frac{1}{p!} (F_{(p)})^{a_1 a_2 \ldots a_p} (F_{(p)})_{a_1 a_2 \ldots a_p}$ (also for $H_{(3)}$). The form-fields $F_{(p)}$ are the gauge invariant field strengths of the potentials $C_{(p-1)}$ (eventually shifted by $H_{(3)}$). Note that the potentials $C_{(p)}$ are not well-defined in the presence of electric and magnetic sources on the D-branes. The self-duality of $F_{(5)}$, i.e. $F_{(5)}=*F_{(5)}$, is not derivable from this action but needs to be imposed by hand. The equations of motion that extremize this action are Einstein's equations for the metric, the equation of motion for the dilaton, and the equations of motion for the form-fields (which are derived and stated in full in Ref. \onlinecite{benini2008}).

\subsection{Ansatze}
For the metric, we concentrate on the convenient ansatz\\
\begin{widetext}
\be
\label{setupmetricansatz}  ds^2 =h(\sigma)^{-1/2}[-b(\sigma) dt^2 + d\vec{x}_3^2] +h(\sigma)^{1/2}\left[b(\sigma) S(\sigma)^8 F(\sigma)^2 d\sigma^2+ S(\sigma)^2 ds_{KE}^2 + F(\sigma)^2 (d\tau + A_{KE})^2\right].
 \ee
 \end{widetext}
Our radial variable is $\sigma$ (with dimensions of length$^4$).
The ansatz above is suitable for a charged black hole in a deformed $AdS$ space times a squashed Sasaki-Einstein manifold.
The squashing is produced by the backreaction of the D7-branes trivially embedded in the geometry (i.e. corresponding to massless flavors).\\

Having the metric functions $b,h,S,F$  depend only on a radial variable $\sigma$ is possible due to the smearing procedure outlined above. In addition, we also have the dilaton $\Phi(\sigma)$ which we also take to depend only on $\sigma$.\\

As we saw above, to introduce a finite baryon density in the dual field theory, we use the following ansatz for the gauge field on the D7-branes:
\be A = A_t(\sigma) dt \, .\ee
In addition, we set $B = H_{(3)} = 0$ throughout. This also implies that $\mathcal{F} = 2\pi \alpha' A_t' d\sigma \wedge dt$ and $\mathcal{F}\wedge {\cal F}=0$. The equations of motion and Bianchi identities which determine the Ramond-Ramond form-fields then become
\bea
\label{setup:F1EOM} d(e^{2\Phi} * F_{(1)}) & = & 0\, ,\\
\label{setup:F3EOM} d(e^{\Phi} * F_{(3)}) & = & 0\, ,\\
\label{setup:F5EOM} d(*F_{(5)}) & = & d F_{(5)} = 0\, ,\\
\label{setup:H3EOM} d(e^{-\Phi} * H_{(3)}) &=& 0  = e^{\Phi} F_{(1)} \wedge * F_{(3)} - F_{(5)} \wedge F_{(3)}\\
\nn && + e^{\Phi} \frac{\delta}{\delta \mathcal{F}} \sqrt{-\det (\hat{G} + e^{-\Phi/2}\mathcal{F})} \delta^{(2)}(D7)\, ,\\
\label{setup:F1bianchi}d F_{(1)} & = & -\Omega_2\, ,\\
\label{setup:F3bianchi}d F_{(3)} & = & - \mathcal{F}\wedge \Omega_2\, .
\eea
Note that in (\ref{setup:H3EOM}), the final term is not explicitly written as a form, following Ref. \onlinecite{benini2008}.
Finally, the equation of motion for the worldvolume gauge field $A_t$ coincides  precisely with the derivative  with respect to $\sigma$ of the right hand side of  (\ref{setup:H3EOM}). Hence with our ansatz  having $B=0$ the equation of motion for $A_t$ is implied by the equation of motion for $H_{(3)}$.\\

To solve these equations of motion and Bianchi identities, it suffices for $F_{(1)}$ and $F_{(5)}$ to assume the ansatze:
\bea
F_{(1)} & = & Q_f (d\tau + A_{KE})\, ,\\
F_{(5)} & = & Q_c (1+ *) \epsilon (X_5)\, ,
\eea
where $\epsilon (X_5)$ is the volume form of $X_5$ and we have introduced the constant $Q_c$ through
\be
N_c = Q_c \frac{\mathrm{Vol}(X_5)}{(2\pi)^4 g_s \alpha'^2}\, .
\ee
Note that at finite temperature, we no longer have supersymmetry to motivate the D7-brane embedding as in (\ref{D3D7:smearDBI}); however, it is checked in Refs. \onlinecite{D3D7QGP}, \onlinecite{Bigazzi:2011it} that the trivial embedding is still consistent for the D7-branes for massless flavors. We will only consider massless flavors in what follows, see Ref. \onlinecite{D3D7QGP} for an example of equations concerning massive flavors at finite temperature.\\

Finally, we are left with the form-field $F_{(3)}$. Motivated by the discussion above, we expect that introducing a non-zero field $A_t$ on the brane will imply that $F_{(3)}$ includes a non-zero component $F_{123} dx^1\wedge dx^2\wedge dx^3$, where we postulate $F_{123}$ to be a constant. However, this is not the end of the story; an analysis of the equations of motion for the form-fields shows us that consistency requires the introduction of additional components of $F_{(3)}$. We can motivate our choice by noticing that there is a WZ term $C_{(6)}\wedge\mathcal{F}$ in the action (\ref{setup:flavoraction}); as soon as $C_{(6)}$ and $\mathcal{F}$ are non-zero, this term must be taken into account. Let us take the following ansatz for the $C_{(6)}$ potential:
\be
C_{(6)} = J(\sigma) dx^{1-3}\wedge (d\tau+A_{KE}) \wedge \Omega_2\, .
\ee
This term gives us two extra terms in our ansatz for $F_{(3)}$ (through $F_{(7)} = dC_{(6)}$ and $F_{(3)} = -e^{-\Phi} * F_{(7)}$). Our full ansatz for $F_{(3)}$ then contains three terms in total\\
\begin{widetext}
\be
 F_{(3)} = F_{123} dx^1\wedge dx^2 \wedge dx^3 - \frac{J' e^{-\Phi}}{S^4 F^2}\,  dt\wedge \Omega_2 + 8 Q_f J e^{-\Phi} b F^2\,  dt\wedge d\sigma\wedge (d\tau + A_{KE})\, .
\ee
\end{widetext}

\subsection{Equations of Motion}
Once we have specified our ansatz, we are left with the metric function $b(\sigma), h(\sigma), F(\sigma), S(\sigma)$ as well as the form-field function $J(\sigma)$, the gauge field $A_t(\sigma)$ and the dilaton $\Phi(\sigma)$; these will be uniquely determined by the equations of motion and the constant $F_{123}$, which is proportional to the baryon density as we will see.\\

First of all, $A_t(\sigma)$ and $J(\sigma)$ are determined in terms of the metric functions by the two equations (\ref{setup:H3EOM}) and (\ref{setup:F3bianchi}), which give us
\bea
\label{setup:Ateq}
2 \pi A_t'(\sigma)  & = &  \frac{(Q_c F_{123} + 8Q_f^2 J) b S^4 F}{\sqrt{16Q_f^2 S^4 F^2 + e^{-\Phi}(Q_c F_{123} + 8Q_f^2 J)^2}} \nonumber\\
& = & -8 J \left(e^{-\Phi} b F^2\right) + \frac{d}{d\sigma}\left[\frac{e^{-\Phi} J'}{F^2S^4}\right].\eea

Then, we need the equations of motion for the metric functions and the dilaton. We can obtain them by calculating the left- and right-hand sides of the Einstein equations and equating them, or by directly inserting our ansatze into the action (\ref{setup:theaction}), (\ref{setup:flavoraction}). In the latter case, we are left with an effective one-dimensional action:\cite{Bigazzi:2011it}
\be\label{azioneunodi}
S = \frac{\mathrm{Vol}(X_5) V_{1,3}}{2\kappa_{10}^2} \int L_{1d} d\sigma\, ,
\ee
where $V_{1,3}$ denotes the (infinite) volume integral over the Minkowski coordinates, and\\
\begin{widetext}
\bea
\nn L_{1d} & = & -\frac12 (\log'h)^2 + 12(\log'S)^2+8\log'F\log'S-\frac12\Phi'^2 +\frac{\log'b}{2}\left(\log'h + 8\log'S + 2\log'F\right) - 4Q_f^2 \frac{J'^2}{F^2S^4}\\
\nn && -\frac{bQ_c^2}{2h^2}-4bF^4S^4 + 24bF^2S^6 - \frac12 F_{123}^2 e^{\Phi}bh^2F^2 S^8 - \frac12 Q_f^2 e^{2\Phi} b S^8-4 e^{\Phi/2} F Q_f S^2\sqrt{-(2\pi\alpha' A_t')^2 + e^{\Phi} b^2 F^2 S^8}\\
\label{setup:1DLagr} &&  - 32Q_f^2 e^{-\Phi} b F^2 J^2 - 8 Q_f^2 (2\pi\alpha'A_t')J \, .
\label{1Dlagrangian}
\eea
\end{widetext}
Since $A_t$ only enters the Lagrangian with its first derivative, it must be connected to a constant of motion. In fact, this constant must be set as follows:
\be
\frac{\delta L_{1d}}{\delta A_t'} = 2\pi\alpha' Q_c F_{123}\, ,
\ee
as this gives us exactly equation (\ref{setup:Ateq}). In fact, we can use (\ref{setup:Ateq}) to eliminate $A_t'$ in favor of $F_{123}$, obtaining the Lagrangian
\be \label{setup:canonicalLagrangian}
\tilde{L}_{1d} = L_{1d} - \left.\frac{\delta L_{1d}}{\delta A_t'} A_t'\right|_{A_t'=A_t'(F_{123})}\, .
\ee
The Euler-Lagrange equations from (\ref{setup:canonicalLagrangian}) then give us the relevant equations of motion and obviously coincide with those obtained directly from the Einstein equations.\cite{Bigazzi:2011it} Finally, the equations of motion as derived from (\ref{setup:canonicalLagrangian}) must be supplemented by the so-called zero-energy condition $H = 0$, with
\be
H = -L_{1d} + \sum_i \psi_i' \frac{\delta L_{1d}}{\delta \psi_i'}\, , \quad \psi_i=\{b,h,F,S,\Phi,A_t,J\}\, ,
\ee
which is a first-order differential equation in the unknown functions.\\

\begin{widetext}
 In any case, the end result is the following set of equations of motion (also restating those determining $A_t$ and $J$ from above)\\
\bea
(\log b)'' & = & b e^{\Phi} F^2 F_{123}^2 h^2 S^8 + \frac{4 (2\pi\alpha'A_t')^2 e^{\Phi/2} F Q_f S^2}{\sqrt{-(2\pi\alpha'A_t')^2 + b^2 e^{\Phi} F^2 S^8}} + 64 b e^{-\Phi} F^2 Q_f^2 J^2
\label{setup:EOMb} + \frac{8e^{-\Phi} Q_f^2 J'^2}{F^2 S^4}\, ,\\
(\log h)'' & = &  -\frac{b Q_c^2}{h^2} + \frac32 b e^{\Phi} F^2 F_{123}^2 h^2 S^8 + \frac{2 (2\pi\alpha'A_t')^2 e^{\Phi/2} F Q_f S^2}{\sqrt{-(2\pi\alpha'A_t')^2 + b^2 e^{\Phi} F^2 S^8}} + 32 b e^{-\Phi} F^2 Q_f^2 J^2
 \label{setup:EOMh}  + \frac{4 e^{-\Phi} Q_f^2 J'^2}{F^2 S^4}\, ,\\
(\log S)'' & = & -2 b F^4 S^4 + 6 b F^2 S^6 - \frac14 b e^{\Phi} F^2 F_{123}^2 h^2 S^8 - \frac{b^2 e^{3\Phi/2} F^3 Q_f S^{10}}{\sqrt{-(2\pi\alpha'A_t')^2 + b^2 e^{\Phi} F^2 S^8}}
  \label{setup:EOMS}- 16 b e^{-\Phi} F^2 Q_f^2 J^2\, ,\\
(\log F)'' & =& 4 b F^4 S^4 -\frac14 b e^{\Phi} F^2 F_{123}^2 h^2 S^8 - \frac12 b e^{2\Phi} Q_f^2 S^8 - \frac{(2\pi\alpha'A_t')^2 e^{\Phi/2} F Q_f S^2}{\sqrt{-(2\pi\alpha'A_t')^2 + b^2 e^{\Phi} F^2 S^8}}
\label{setup:EOMF} + 16 b e^{-\Phi} F^2 Q_f^2 J^2\nonumber \\
&& - \frac{2e^{-\Phi} Q_f^2 J'^2}{F^2 S^4}\, ,\\
\Phi'' & = & \frac12 b e^{\Phi} F^2 F_{123}^2 h^2 S^8 + b e^{2\Phi} Q_f^2 S^8 + \frac{2 b^2 e^{3\Phi/2} F^3 Q_f S^{10}}{\sqrt{-(2\pi\alpha'A_t')^2 + b^2 e^{\Phi} F^2 S^8}} + 2 e^{\Phi/2} F Q_f S^2 \sqrt{-(2\pi\alpha'A_t')^2 + b^2 e^{\Phi} F^2 S^8} \nonumber\\
&& - 32 b e^{-\Phi} F^2 Q_f^2 J^2
\label{setup:EOMphi}  - \frac{4 e^{-\Phi} Q_f^2 J'^2}{F^2 S^4}\, ,\\
\left[\frac{e^{-\Phi} J'}{F^2S^4}\right]' &= & \frac{(Q_c F_{123} + 8Q_f^2 J) b S^4 F}{\sqrt{16Q_f^2 S^4 F^2 + e^{-\Phi}(Q_c F_{123}+ 8Q_f^2 J)^2}}
\label{setup:EOMJ}  + 8 J \left(e^{-\Phi} b F^2\right)\, ,\\
\label{setup:EOMAt} 2 \pi A_t'(\sigma)  & = &  \frac{(Q_c F_{123} + 8Q_f^2 J) b S^4 F}{\sqrt{16Q_f^2 S^4 F^2 + e^{-\Phi}(Q_c F_{123} + 8Q_f^2 J)^2}}\, ,
\eea
together with the first-order constraint
\bea
\nn 0 & = & -\frac12\log'h\log'b + \frac12 (\log'h)^2-12(\log'S)^2-4\log'b\log'S - \log'b\log'F - 8\log'F\log'S + \frac12\Phi'^2\\
\nn && -\frac{bQ_c^2}{2h^2} - 4bF^4S^4 + 24bF^2S^6 - \frac12 F_{123}^2e^{\Phi}bh^2F^2S^8 - \frac12be^{2\Phi}Q_f^2 S^8
 -\frac{4b^2 e^{3\Phi/2}F^3 Q_f S^{10}}{\sqrt{-(2\pi\alpha'A_t')^2+b^2e^{\Phi}F^2S^8}} \\
&& - 32be^{-\Phi}F^2Q_f^2J^2 + \frac{4e^{-\Phi}Q_f^2J'^2}{F^2S^4}\, .
 \label{apconstraint}
\eea
\end{widetext}
Now that we have obtained the equations of motion, we can finally proceed in order to find their solutions.

\section{The Perturbative Solution}
\label{sec:solution}

In order to find a complete solution of the equations of motion, we should in principle proceed by numerical integration. Fixing the asymptotic behaviors one should glue them numerically so as to determine the integration constants.
This kind of analysis is especially difficult in the case at hand due to the presence of a Landau pole, and stability of the numerical integration becomes a very relevant issue.\\

The appearance of a  Landau pole is expected on  physical grounds  as we are perturbing conformal field theories, as e.g. ${\cal N}=4$ supersymmetric  Yang-Mills, with the addition of fundamental matter. The one-loop beta function will be positive and proportional to $N_f$ and the running coupling constant will have a divergence at some UV scale $\Lambda_{UV}$. In the holographic dual it will be seen as a divergence of the dilaton field.
As is the case for QED, the meaning of this is that we are dealing with an IR effective theory,  and a suitable completion should replace it in the UV before the Landau pole is reached.\\

To gain information about the full solution, a clever approach is to follow a procedure analogous to the one considered in Ref.
\onlinecite{Gubser:2001ri} to construct an approximate black hole solution of the Klebanov-Tseytlin model. Given the fact that, for $M$ the number of fractional branes and $N$ that of regular D3-branes, the exact black hole is known when $M=0$,  Ref. \onlinecite{Gubser:2001ri} considered a perturbative expansion with parameter $M^2/N \ll 1$. Already at first order, such perturbative analysis gave notable results.
The same strategy, supplemented by a numerical analysis, has been applied in Ref. \onlinecite{Herzog:2009gd}, where a solution at finite charge density for \emph{di}-baryons has been discussed.\\

In the flavored setup, the natural parameter for a perturbative expansion is given by the effective coupling $g_{FT}^2 N_f$ which weighs the vacuum polarization effects due to the dynamical flavors. In the unquenched smeared
D3-D7 model, the supergravity plus DBI solution is reliable in the Veneziano limit $N_f, N_c \to \infty$ with $\lambda N_f/N_c =$ constant, provided the 't Hooft coupling $\lambda = g_{FT}^2 N_c$ is large. This in turn implies that $N_f/N_c\ll1$ and
the effective coupling $g_{FT}^2 N_f \sim \lambda N_f/N_c$ can be of order one.
Inspired by Ref. \onlinecite{Gubser:2001ri} we will now consider the perturbative regime in the Veneziano parameter $\epsilon\sim \lambda N_f/N_c\ll 1$.\\

In this section we will derive an analytical perturbative solution at finite temperature which has a nontrivial backreaction of the flavor degrees of freedom, as well as
a nonzero baryon chemical potential. The price of having an analytic solution is
that we must keep the deformation  small, so that we can perform a perturbative expansion around the original unflavored theory.
\\

 In a first approach we expect roughly three independent parameters, which we will name $r_h, \epsilon$ and $\delta$.
The first one is related to the temperature, $r_h\sim T $. For $\epsilon=\delta=0$ it will parametrize the usual $AdS_5$ black hole solution with  $T = r_h/(\pi\sqrt{\lambda} \alpha')$. The second one, $\epsilon \sim Q_f$, is a measure of the backreaction of the flavor degrees of freedom and, in fact,  turns out to be related to the flavor-loop counting parameter. Finally, related to the baryon density
we will have another parameter $\delta \sim A_t'$ .
\\

Concerning $\epsilon$, it turns out that the relevant flavor counting parameter is scale dependent.
The best way to think of it is to notice that, in the equations of motion (\ref{setup:EOMb})--(\ref{setup:EOMAt}), at zero baryon density ($n_b=0 \Rightarrow J=A_t=F_{123}=0$), $Q_f$ and the dilaton $\Phi$ always appear in the combination $Q_f e^\Phi$.
Hence, consider a  natural split of the dilaton field at an arbitrary ``anchoring" scale $\Lambda_* \sim  \sigma_*$, $\Phi(\sigma) = \Phi_{*} + \phi(\sigma)$ where $\phi(\sigma_*) = 0$. Then
 $\epsilon_*$  at that scale is defined by
\be
\epsilon_* = Q_f e^{\Phi_*}  = \frac{{\mathrm{Vol}}(X_3)}{16\pi {\mathrm{Vol}}(X_5)}\lambda_* \frac{N_f}{N_c} \label{epslambda1} \, ,
\ee
where $\lambda_*=4\pi g_s e^{\Phi_*} N_c$ and we have made use of (\ref{flavdens}).  Hence $\epsilon_*$   weighs the backreaction of the
D7-branes and, in fact, can be read as a flavor-loop counting
parameter in the dual field theory.
Keeping $\epsilon_*$ small is tantamount to 
imposing a large separation of the scale $\sigma_*$ and the dangerous Landau pole $\sigma_{LP}$.\cite{D3D7QGP}

In order to derive an analytical perturbative solution we will take  both $\epsilon_*$ and $\delta$ to be much smaller than
one. All the functions in the ansatz will be expanded in  power series of $\epsilon_*^m \delta^n$. To get started   let us introduce
the following leading scaling behaviors
 \be
  \quad F_{123} =  \epsilon_*^{1/2} \delta   \frac{\sqrt{Q_f}}{Q_c}\, ,~~~
  \quad  J(\sigma) = \epsilon_*^{3/2} \delta \frac{\tilde J(\sigma) }{Q_f^{3/2}}  \label{scalingdelta}\, .
\ee
The reason behind this choice goes as follows:
inserting these expressions into
 (\ref{setup:EOMb})--(\ref{setup:EOMAt}) and recalling  the dilaton splitting
as $\Phi(\sigma)=\Phi_* + \phi(\sigma)$, with $\phi(\sigma_*)=0$,
one readily arrives at the following system of equations\\
\begin{widetext}
\bea
(\log b)'' &=& 4\, {\epsilon_*\delta^2 \, }\, \frac{X}{Y} + 64\, \epsilon_*^2\delta^2 \, e^{-\phi} \, b F^2  \tilde J^2 + 8\, \epsilon_*^2\delta^2 \, e^{-\phi} \,  \frac{ \tilde J'^2}{F^2 S^4} + \epsilon_*^2\delta^2 \,  \,Z\, , \label{eomb} \\
(\log h)'' &=& - Q_c^2 \frac{b}{h^2} + 2\, \epsilon_*\delta^2 \,  \, \frac{X}{Y} + 32 \, \epsilon_*^2\delta^2 \, e^{-\phi} \,b F^2 \tilde J^2 + 4\, \epsilon_*^2\delta^2 \, e^{-\phi} \,\frac{\tilde J'^2}{F^2 S^4} + \epsilon_*^2\delta^2 \,  \,\frac{3}{2}Z \, ,\label{eomh}\\
(\log S)'' &=& - 2 b F^4 S^4 + 6 b F^2 S^6 - \epsilon_* \, e^{3\phi/2}\frac{b^2 F^3 S^{10}}{Y} - 16\, \epsilon_*^2\delta^2 \, e^{-\phi} \, b F^2 \tilde J^2  - \epsilon_*^2\delta^2 \,  \,\frac{1}{4}Z \, ,     \label{eomS} \\
(\log F)'' &=& 4 b F^4 S^4 -\frac{1}{2}\epsilon_*^2 e^{2\phi}\, b S^8 - \epsilon_*\, \delta^2 \, \, \frac{X}{Y} + 16\, \epsilon_*^2\delta^2 \, e^{-\phi} \, b F^2 \tilde J^2 - 2 \, \epsilon_*^2\delta^2 \, e^{-\phi} \,\frac{\tilde J'^2}{F^2 S^4}    \label{eomF} - \epsilon_*^2\delta^2 \,  \,\frac{1}{4}Z\, , \\
(\phi)'' &=&  \epsilon_*^2 e^{2\phi} b S^8 +2 \epsilon_* e^{3\phi/2} \frac{b^2 F^3 S^{10}}{Y} + 2  \epsilon_* e^{\phi/2} F S^2 Y -32 \epsilon_*^2\delta^2 e^{-\phi} b F^2\tilde J^2  \label{eomphi}
 - 4  \epsilon_*^2\delta^2 e^{-\phi} \frac{\tilde J'^2}{F^2 S^4}+ \epsilon_*^2\delta^2 \frac{1}{2}Z\, ,\\
\left[ \frac{e^{-\phi}\tilde J'}{S^4 F^2} \right]' &=& \frac{( 1+8\epsilon_* \tilde J)bFS^4}{\sqrt{16 F^2 S^4+ \delta^2e^{-\phi}(1+8\epsilon_* \tilde J)^2    }}
+8e^{-\phi}b  F^2 \tilde J \, .
\label{eomj}
\eea
The  constraint equation (\ref{apconstraint}) reads\\
\bea
 0 & = & - \frac{1}{2} \log' h \log' b + \frac12 (\log' h)^2 -
12(\log'S)^2 - 4\log'b \log'S  - \log'b \log'F - 8 \log'F \log'S + \frac12 \phi'^2 \label{constraint} \\
 & & - \frac{b Q_c^2}{2 h^2} - 4 b F^4 S^4 + 24 b F^2 S^6 -\epsilon_* \frac{4 e^{3\phi/2}b^2 F^3  S^{10}}{Y}  - \epsilon_*^2 \frac12 e^{2\phi} b  S^8 +\epsilon_*^2 \delta^2\left(- 32   b e^{-\phi} F^2  \tilde J^2 +  \frac{4 e^{-\phi}   \tilde J'^2}{F^2 S^4}-\frac12 Z\right) . \,
\nonumber
\eea
\end{widetext}

In these equations  $X$, $Y$, and $Z$ are shorthand for\\
\bea
X &=&  \frac{(1+8 \epsilon_* \tilde J)^2   e^{\phi/2}b^2 F^3S^{10}}{ 16   F^2 S^4+\delta^2 e^{-\phi}(1+8 \epsilon_* \tilde J)^2     }\, ,
 \\
Y &=& \sqrt{b^2 e^\phi F^2 S^8- \frac{\delta^2(1+8  \epsilon_* \tilde J)^2b^2F^2S^8}{  16   F^2 S^4+ \delta^2 e^{-\phi}(1+8 \epsilon_* \tilde J)^2  } }
 \\
Z &=& \frac{e^{\phi} \, b h^2 F^2 S^8}{Q_c^2} \label{xyzscaling}\, .
\eea

The system (\ref{eomb})-(\ref{eomj})  allows for a
systematic expansion of all the functions in powers series of
$\epsilon_*$ and $\delta^2$. This is essentially the main effect of the scaling relations (\ref{scalingdelta}).
Once all the functions have been solved for, the worldvolume gauge
field can be obtained from the following relation \be 2\pi\alpha'
A_t' =\delta e^{\Phi_*/2} \frac{(1+8\epsilon_* \tilde
J)bFS^4}{\sqrt{16 F^2 S^4+ \delta^2 e^{-\phi}(1+8\epsilon_* \tilde
J)^2 }} \label{gfieldprof} \, , \ee which is already first order in
$\delta$. From this expression it is fairly obvious  that
$J(\sigma)$ is a higher order effect in $\epsilon_*$ and hence arises as backreaction of the baryon density
onto the flavor branes.\\

Actually, it is convenient to introduce a
dimensionless parameter $\tilde\delta$  defined as \be \tilde\delta=\frac{\delta}{4 r_h^3}\,, \ee where  the factor
$4$ is introduced in order to make it  precisely $\tilde\delta =\tilde d$
of Ref. \onlinecite{hep-th/0611099}. We keep the Greek symbol,
however, in order to stress that our parameter is going to be
perturbatively small.\\

The strategy now is to set up a perturbative ansatz in both $\epsilon_*$ and $\tilde\delta$, by means of which, all functions
admit an expansion like
\be
\Phi(r) = \sum_{i,j\geq 0} \Phi_{ij}(r) \epsilon_*^i \tilde\delta^{2j}\, .
\ee

The  reader may have noticed that we have changed radial coordinate $\sigma\to r$. This is done after integrating the equations of motion in $\sigma$ and the
new radial  coordinate ``gauge" is chosen to make the $d\vec{x}_3^2$ component of the metric equal to
$r^2/R^2$, just the same as in the unflavored  solution. This allows for an easy comparison of the
two situations.\\

Equations (\ref{eomb})-(\ref{eomj}) are solved for order by order in $\epsilon_*$, $\tilde\delta$.
At each order, two conditions must be imposed in order to fix the two integration constants that appear in each function. One of them will be regularity at the horizon
$r=r_h$. The other one will be matching the supersymmetric solution found in Ref. \onlinecite{Benini:2006hh} at some UV energy scale $r=r_s$.
In summary we have three energy scales: the horizon radius $r_h$, the dilaton anchoring point $r_*$, and the UV cutoff scale $r_s$.
$r_*$ and $r_s$ are arbitrary, and in Ref. \onlinecite{D3D7QGP} they were fixed to $r_*=r_s$.\\

A solution containing the three parameters is unwieldy, and we sketch here just an example up to ${\cal O}(\epsilon_*)$\\

\begin{widetext}
\bea
\phi(r) &=&  \epsilon_* \log\frac{r}{r_*} + \frac{\epsilon_*^2}{72} \left[ \left( 1+ 6\log\frac{r}{r_*}\right)^2  - \frac{2r^4-r_h^4 + 2(r_s^4-r_*^4)}{2 r_s^4-r_h^4}
+\frac{9}{2} \left(  \hbox{Li}_2\left(1 - \frac{r_h^4}{r^4}\right) - \hbox{Li}_2\left(1- \frac{r_h^4}{r_*^4}\right)\right)
\right]+...\, ~~~\label{rswrstar}
\eea
\end{widetext}
which naturally satisfies $\phi(r_*)=0$. In Ref. \onlinecite{D3D7QGP} a solution to order $\epsilon_*^2$ was presented in equation (2.22) with $r_*=r_s$. Namely, the scale where the  't Hooft coupling is defined and
the one where the thermal solution matches the supersymmetric one coincide.
The solution is well defined in the range $r_h<r<r_*$.
If one is interested in IR quantities, like thermodynamic properties, or transport coefficients, it is more sensible to work with a solution where all  the running parameters  are defined at the relevant scale, e.g. with $\epsilon_h = Q_f e^{\Phi(r_h)}$.
The solution with $r_*=r_h$ and $r_s\to \infty$ was used in section 3 of Ref. \onlinecite{D3D7QGP} as well as in eq. (2.7) of  Ref. \onlinecite{hydro1}.
In Ref.  \onlinecite{Bigazzi:2011it} we presented an extended version of the  solution which contained corrections up to order $\tilde\delta^2$. In that paper, rather than the metric coefficients, we gave the expression for the
functions in the ansatz, $F$, $S$, etc. We will provide here a final form for the metric, dilaton and $A_t$ potential.
Namely, writing the line element as\\
\begin{widetext}
\bea
ds^2 \! &=& G_{tt} dt^2 + G_{xx}  d\vec x_3^2 + G_{rr} dr^2 + G_{_{\!K\! E}} ds_{_{\!K\! E}}^2 + G_{\tau\tau} (d \tau + A_{_{\!K\! E}})^2\, ,
\label{metricansatz}
\eea
we find
\bea
G_{tt} (r)& =& - \frac{r^2}{R^2}\left(1-\frac{r_h^4}{r^4}\right)  + \epsilon_h\tilde \delta^2
 \frac{r_h^2(2  r^4  - r_h^4) + r^2 r_h^4(\log 2 - 1) + r^2(r_h^4-2r^4)\log\left(1+\frac{r_h^2}{r^2}\right) }{2 r^4 R^2} \nonumber\\
 &&  ~+\,\epsilon_h^2\tilde \delta^2 \frac{-34 r^4r_h^2 +5 r_h^6+r^2 r_h^4(29-17\log 2)+17 r^2(2 r^4-r_h^4)\log\left(1 + \frac{r_h^2}{r^2}\right)}{24 r^4 R^2}+... \, ,\label{solgtt}
 \\
G_{xx}(r) &=& \frac{r^2}{R^2} \, ,\label{solgxx}\\
G_{rr}(r) &=&\frac{R^2}{r^2}\frac{1}{\displaystyle 1-\frac{r_h^4}{r^4}}\left\{\rule{0mm}{7mm} 1 +  \right. \nn\\
&&\left. +\,  \epsilon_h \left[\frac{1}{4}+  \tilde\delta^2 \left( \frac{-6r^8 r_h^2 + 3 r^6 r_h^4 + 7 r^4 r_h^6 - r_h^{10}+r^2 r_h^8(\log 4 - 3) + r^2(6r^8-9r^4r_h^4+r_h^8)\log\left(1+\frac{r_h^2}{r^2}\right)}{4r_h^4 r^2(r^4-r_h^4)} \right) \right]
\right. \nonumber\\
&& +\, \epsilon_h^2 \left[\rule{0mm}{7mm} \frac{11-24\log\displaystyle \frac{r_h}{r}}{96}
 \right. \label{solgrr}\nonumber \\
&& \left. \left. - ~ \tilde \delta^2 \frac{-42 r^8r_h^2 + 21 r^6r_h^4+49r^4 r_h^6+5 r_h^{10}+r^2r_h^8\left(-33+14 \log 2 \right)+7r^2(6r^8-9r^4r_h^4+r_h^8)\log\left(1 + \frac{r_h^2}{r^2}\right) }{24 r_h^4(r^4 - r_h^4)}\right]
\right\}\nonumber\\
&& +... \, , \\
 G_{_{\!K\! E}}(r) &=& R^2\Biggl\{ 1   + \epsilon_h \frac{1}{12} + \epsilon_h^2 \left(\frac{ 5 - 24 \log \displaystyle\frac{r_h}{r}}{288} \right)  + \tilde \delta^2 \left( \frac{\epsilon_h}{20}+\frac{\epsilon_h^2}{15}\right)\times \nonumber\\
&& ~~~ \times \frac{1}{r^2r_h^4} \left[  \frac{1}{2} r_h^4 r^2 G(r)
+ r^2\left(  2r^2 r_h^2 - 3 r_h^4+3(-2r^4 + r_h^4)\log \left(1 + \frac{r_h^2}{r^2} \right)\right)\right] \Bigg\} +...\, , \label{solgke} \\
G_{\tau\tau} (r)&=& R^2\Bigg\{ 1   - \epsilon_h \frac{1}{12} + \epsilon_h^2 \left(\frac{ 3 +8 \log\displaystyle\frac{r_h}{r}}{96} \right)  + \tilde \delta^2 \left( \frac{\epsilon_h}{20} + \frac{3\epsilon_h^2}{40}\right)\times \nonumber\\
&& ~~~ \times \frac{1}{r^2r_h^4} \left[ - 2r_h^4 r^2 G(r)
+ r_h^2\left(  22r^4  - 3r^2 r_h^2 - 5 r_h^4\right) + 3r^2(-2r^4+r_h^4)\log \left(1 + \frac{r_h^2}{r^2} \right)\right] \Bigg\} +...\, , \label{solgtautau}\\
\phi(r) &=& \rule{0mm}{10mm}  \epsilon_h \log\frac{r}{r_h} + \epsilon_h^2\left[\rule{0mm}{7mm} \frac{1}{6}\left( 1 + 3\log\frac{r}{r_h}\right)\log\left(\frac{r}{r_h}\right) +
\right.
 \label{solphi} \\
&&\left. \frac{1}{120}\tilde\delta^2 \left( 41 - 2\pi - \frac{26 r^2}{r_h^2} -\frac{15 r_h^2}{r^2} +G(r) - 29\log 2 +
\left( 11 + \frac{18 r^4}{r_h^4}\right) \log\left( 1 + \frac{r_h^2}{r^2}\right) + \frac{1}{16}\hbox{Li}_2 \left(1 - \frac{r_h^4}{r^4}\right)    \right) \rule{0mm}{7mm}\right]\, ,
\nn \\
A_t'(r) &=& \rule{0mm}{10mm} \frac{ e^{\Phi_h} }{2\pi\alpha'  }\frac{r_h^3}{r^3}\tilde\delta\left( 1 - \frac{\epsilon_h}{6} - \frac{\epsilon_h^2}{288}\left(1 - 24\log\frac{r}{r_h}\right) + ...\right) \, ,
 \label{solgatp}\\
  \tilde J(r) & = &\rule{0mm}{10mm}  - \frac{r_h^3}{8} + ...\,,   \label{solj}
\eea
\end{widetext}
where dots stand for higher orders $\epsilon_h^{>2}\tilde\delta^{>2}$.  $G(r)=2\pi
\frac{r_h^6}{r^6}\,{}_2F_{1}\left(\frac{3}{2},\frac{3}{2},1,1-\frac{r_h^4}{r^4}\right)$
is a hypergeometric function and $ \hbox{Li}_2(u)\equiv \sum_{n=1}^\infty
\frac{u^n}{n^2}$ is a polylogarithmic function.
$R=(Q_c/4)^{1/4}$ is the overall radius of the internal space (and of the $AdS$ factor in the unflavored case).
Notice that $\tilde J$
enters always multiplied by $\epsilon_h^2$ in the equations of motion (\ref{eomb})-(\ref{eomj}), hence only the
leading contribution in $\tilde J$ is relevant in the solution.\\

The  horizon radius $r_h$  satisfies   $G_{tt}(r) = (r-r_h) b_0(r) +{\cal O}(\epsilon_h^3, \tilde\delta^4)$ with $b_0(r_h)\neq 0$. As mentioned before, the radial coordinate $r$ is derived after the
solution has been obtained in $\sigma$, by demanding that $G_{xx}(r)=r^2/R^2$ just as in the unflavored supersymmetric solution.
 Its relation to the coordinate $\sigma$ in which the equations of motion acquire the form given in (\ref{eomb})--(\ref{eomj}) is also perturbative\\
\begin{widetext}
\bea
r &=& \frac{r_h}{\left(1-e^{4 r_h^4 \sigma}\right)^{1/4}} \left[\rule{0mm}{6mm} 1 + \frac{1}{16}\epsilon_h\tilde\delta^2
\left(
2\sqrt{1-e^{4r_h^4\sigma}} - 2 - 2\tanh^{-1} \sqrt{1-e^{4r_h^4\sigma}}+ \rule{0mm}{5mm}
 \right.\right.\nonumber\\
&&\left. \rule{0mm}{6mm}\left.~~~~~~~~~~~~~~~~~~~~~~~~~~~~~~~~~~~~~~~~~~~~~~~~~~~~\rule{0mm}{5mm} + 4 r_h^4\sigma(1+\coth(2r_h^4\sigma))(1-\log 2) + \log 4 - 4 r_h^4\sigma
\right)+ ...\right]\, .
\eea
\end{widetext}

The one above must be understood as an effective IR solution, where the UV has been decoupled (i.e. all terms of the form $r_h/r_s$ have been dropped). It should not be
used for computing physical quantities at energies much higher than the plasma temperature.\\

Let us conclude this section by summarizing the regime of validity of the solution.
Apart from the usual bounds $N_c \gg 1,\ \lambda \gg 1$ needed for the gravity approximation to be reliable, we must have $1 \ll N_f \ll \sqrt{\lambda} $, where the lower bound comes from the requirement of validity of the smeared D7-branes configuration, while the upper bound is due to the use of the Abelian DBI  (see section \ref{sectionvalidity}).
Finally, the perturbative solution is reliable if $\epsilon_h \ll 1,\ \tilde \delta \ll 1$; the first condition allows to push the Landau pole far in the UV and have a predictive small temperature solution.
Note that, as we will show in the next section, $\tilde \delta \sim \mu/\sqrt{\lambda}T$, so the requirement $\tilde \delta \ll 1$ means that the chemical potential $\mu$ must not be parametrically larger than the temperature $T$.

\section{Thermodynamics}\label{section:thermodynamics}

In the previous section we have constructed a family of solutions consisting of black D3-branes and D7-branes smeared along the directions transverse to their worldvolumes. These branes are dressed by a set of fields (scalar, vectorial and higher rank tensor forms), and depend parametrically on the coefficients $\epsilon_h$ and $\tilde \delta$. The size of the event horizon is dictated by these two parameters, and therefore thermodynamic properties associated to the horizon, for example the temperature or the entropy density, are given in terms of them.\\

Given the solution we presented in \eqref{solgtt}-\eqref{solj}, perturbative in $\epsilon_h$ and $\tilde \delta$, we will perform  the thermodynamic analysis up to order $\epsilon_h^2$ and $\tilde \delta^2$, which suffices to catch effects due to the backreaction of the fundamental matter. Consequently, all the results given in this section will have corrections of order ${\cal O}(\epsilon_h^3,\tilde \delta^4)$ which we will not write explicitly. Moreover, we will focus on the near-IR part of the geometry, equivalent to pushing the Landau pole to infinity (i.e. considering the limit $r_s\to\infty$ as commented in the previous section). Thus, the results in this section have to be understood as providing correct answers up to subleading corrections ${\cal O} \left( \frac{r_h^4}{r_s^4} \right) \ll 1$.\\

The position  of the event horizon, $r_h$, is determined by the solution to the equation $b(r_h)=0$. For finite $\tilde \delta$ there are two roots of the blackening function, and in this work we focus on the larger one. As the value of $\tilde \delta$ is increased the two roots approach and, eventually, they are expected to merge, giving rise to an extremal horizon with zero temperature and finite  entropy density. This situation is beyond the regime of validity of our solution, though, and will not be pursued.\\

After Wick rotating time and imposing the absence of conical singularities in the geometry, we obtain the temperature (the inverse of the periodicity of the Euclidean time)\\
\begin{widetext}
\be\label{JT.temperature}
T  = \frac{r_h}{\pi R^2}\left[1- \frac{\epsilon_h}{8} \left(1 + \tilde \delta^2 (2-\log 2) \right) - \frac{13}{384}\epsilon_h^2\left(1 + \tilde \delta^2\frac{2(37 \log 2-26)}{13}  \right)+...\right]  \, .
\ee
\end{widetext}

The entropy, $S$, is given by the Bekenstein-Hawking law and is  proportional to the area of the horizon (an eight-dimensional, fixed Euclidean time, fixed $r=r_h$ hypersurface). As the solution considered here is an extended black brane, this area is infinite due to the integration along the spatial coordinates. To avoid the volume divergence we will consider intrinsic densities, $s=S/\int d^3 x$, and the (finite) entropy density reads
\be\label{JT.entropy}
s=   \frac{\pi^5N_c^2T^3}{2{\mathrm{Vol}}(X_5)} \left[1+ \frac{1}{2}\epsilon_h(1 + \tilde \delta^2) + \frac{7}{24}\epsilon_h^2(1 + \tilde \delta^2)\right] \ ,
\ee
where we have inverted \eqref{JT.temperature} to express the position of the horizon in terms of the temperature and $\mathrm{Vol}(X_5)$ is the volume of the five-dimensional Sasaki-Einstein manifold.

The leading ${\cal O}(\epsilon_h^0 \tilde \delta^0)$ term in expression \eqref{JT.entropy} is the known unflavored result. This will be always the case in the results that  follow. The ${\cal O}(\epsilon_h\, \tilde \delta^0)$ term coincides with the result obtained in the quenched approximation.\cite{mmt} This agreement may seem surprising due to the different strategies followed to derive the result. In the present work the black hole  size increases (with respect to the unflavored setup) due to the presence of fundamental matter in the system. Furthermore, the branes describing the fundamental matter are \emph{distributed} along the  directions transverse to the worldvolume of each individual D7-brane, explicitly breaking the symmetry group $U(N_f)\to U(1)^{N_f}$. On the contrary, in the quenched calculation a set of $N_f\ll N_c$ \emph{coincident} probe D7-branes on the background given by the $N_c$ D3-branes was considered.\cite{mmt}  Considering additivity of the system, the entropy associated to these degrees of freedom was calculated from the free energy and then added up to the entropy associated to the adjoint matter.
A similar discussion holds for the ${\cal O}(\epsilon_h\, \tilde \delta^2)$ term and the results of the quenched approximation in the presence of a finite baryon density or a chemical potential.\cite{hep-th/0611099} Finally,  terms that are higher order  in $\epsilon_h$ or $\tilde \delta$ are  new results.\\

The next quantity that can be derived just with the  metric obtained in the previous section is the ADM energy  of the black hole. This quantity is not defined at the horizon but at the UV cutoff $r_s\to\infty$. Formally it is given by
\be
E_{ADM} = -\frac{1}{\kappa_{10}^2} \int d S_{ab} K^{ab} \ ,
\ee
where $K_{ab}$ is the extrinsic curvature of a hypersurface at constant time and radius. This expression diverges in the $r_s\to\infty$ limit and must be regularized. A useful subtraction scheme is given by considering $E_{ADM} \to E_{ADM} - E_{ADM}^{(0)}$ where $E_{ADM}^{(0)}$ corresponds to the ADM energy of the solution with $T=\tilde \delta=0$. Details of this calculation can be found elsewhere\cite{D3D7QGP} and we quote the final result for the subtracted energy density $\varepsilon=(E_{ADM}-E_{ADM}^{(0)})/\int d^3 x$:
\be\label{JT.subtE}
\varepsilon = \frac{3\pi^5 N_c^2 T^4  }{8{\mathrm{Vol}}(X_5)}\left[ 1+ \frac{1}{2} \epsilon_h \left( 1+2\tilde \delta^2 \right) + \frac{1}{3} \epsilon_h^2\left( 1+\frac74 \tilde \delta^2 \right)  \right]\ .
\ee

Considering only ${\cal O}(\tilde \delta^0)$ terms in \eqref{JT.subtE} the thermodynamic relation $d\varepsilon=T ds$ is satisfied provided
\be\label{JT.gaugerunning}
\frac{\partial \epsilon_h}{\partial T} = \frac{\epsilon_h^2}{T} + {\cal O}(\epsilon_h^3) \ .
\ee
This expression can be derived independently from the fact that the gauge coupling constant $\lambda_h=4\pi g_s e^{\Phi_h}N_c$
is now scale dependent, as the dilaton $\Phi_h = \Phi(r_h)$ runs at first order  in $\epsilon_h$, see (\ref{solphi}).
 This is dual to  the logarithmic  running of the gauge coupling induced by the flavors, $\partial \lambda_h / \partial T = \epsilon_h \lambda_h/T + \cdots$. Actually, the solution for the dilaton at this first order in $\epsilon_h$ is not dependent on $\tilde \delta$, and therefore the relation \eqref{JT.gaugerunning} is valid when ${\cal O}(\tilde \delta^2)$ terms are considered. The inclusion of the $\tilde \delta$ parameter in the discussion leads to the inclusion of a chemical potential/charge density term in the thermodynamic relation,  $d\varepsilon = T ds +\mu \,d n_q$. The determination of $n_q$, the quark density of the system, can be carried out holographically from the action (\ref{azioneunodi}) as
\be\label{JT.nq}
n_q = \frac{\delta S}{\delta F_{t\sigma}} =  \frac{\pi^{5/2}{\mathrm{Vol}}(X_3) N_f N_c\sqrt{\lambda_h} T^3}{16{\mathrm{Vol}}(X_5)^{3/2}}  \tilde \delta\, \left[1 + \frac{3}{8}\epsilon_h\right] \ ,
\ee
where we have used equation \eqref{JT.temperature} and $R^4=Q_c/4$. At leading order this expression coincides with the result obtained in the quenched approximation.\cite{hep-th/0611099}\\

The product of the quark density, $n_q$, with the chemical potential, $\mu$, can be obtained by considering the thermodynamic potentials in the canonical and grand-canonical ensembles, $f=\varepsilon-Ts$ and $\omega=\varepsilon-Ts-\mu n_q$ respectively. These are related to each other by a Legendre transform, and with the Lagrangian \eqref{1Dlagrangian} by an on-shell evaluation, as we will see later. Accordingly, one is able to obtain an expression for $\mu\,n_q$ by evaluating on-shell the term
\be\label{JT.munq}
\mu \, n_q=f-\omega =\frac{1}{\int d^3x} \int \frac{ \delta {\cal L}}{ \delta A_t'} A_t' \Bigg|_{r_s\to\infty}   \ ,
\ee
and with \eqref{JT.nq} we can disentangle the expression for the chemical potential
\be\label{JT.mu}
\mu =  \frac{\pi^{3/2}\sqrt{\lambda_h}T}{4{\mathrm{Vol}}(X_5)^{1/2}}  \, \tilde \delta  \left[ 1+\frac{5}{24}\epsilon_h \right]  = A_t^{(UV)}\left( 1+\frac{\epsilon_h}{4} \right) \ ,
\ee
and once again the leading term was known from the calculation in the 't Hooft limit.\cite{hep-th/0611099}\\

With this, the satisfaction of the relation  $d\varepsilon = T ds +\mu \,d n_q$ at order ${\cal O}(\epsilon_h^2 \tilde \delta^2)$ is guaranteed provided
\be\label{JT.dddTnq}
 \left( \frac{d \, \tilde \delta}{d T}\right)_{n_q} = - \frac{\tilde \delta}{T} \left[ 3 +\frac{\epsilon_h}{2}  + {\cal O}(\epsilon_h^2) \right] \ ,
\ee
i.e., that we work at fixed $n_q$, such that the variation of $\tilde \delta$ in \eqref{JT.nq} is compensated by the variation of the temperature and the gauge coupling.  This in turn is a consequence of the fact that we are checking the thermodynamic relation in the canonical ensemble. To check it in the grand-canonical one  we would need to compare the ADM energy not to the $T=\delta=0$ case, but to the extremal one in which $b(r_h)=b'(r_h)=0$ at fixed chemical potential $\mu_0$. In this case the thermodynamic relation to satisfy is $d(\varepsilon-\varepsilon_{ext})=T ds+(\mu-\mu_0)dn_q$, however, this subtraction scheme is beyond the range of validity of our solution and we do not pursue it. As a result, in the grand-canonical ensemble the  variation with the temperature at fixed chemical potential is obtained from equation \eqref{JT.mu}, resulting in
\be\label{JT.dddTmu}
 \left( \frac{d \, \tilde \delta}{d T}\right)_\mu = - \frac{\tilde \delta}{T} \left[ 1 +\frac{\epsilon_h}{2}  + {\cal O}(\epsilon_h^2) \right]\, \ .
\ee

In the previous analysis we have used the definition of the free energy in the canonical and grand-canonical ensembles. The right-hand side of these definitions have  been discussed in some extent now, and a relation of the left-hand side with the on-shell evaluation of the action was outlined. Next, we proceed  to check that these two calculations agree with each other.\\

The Lagrangian $\tilde L_{1d}$ in \eqref{setup:canonicalLagrangian}, depending  on $F_{123}\sim \tilde \delta$, is the appropriate one for computing the free energy in the canonical ensemble, where the natural variable to consider is the quark density. The action derived from this Lagrangian has to be supplemented with a Gibbons-Hawking term to deal with a well-posed variational problem. Despite the addition of this term, the evaluation of the on-shell euclidean action presents divergences in the UV ($r_s\to\infty$). These divergences can be tamed with the inclusion of an appropriate series of counterterms or by the subtraction of a suitable reference background. As we did in \eqref{JT.subtE} for the ADM energy of the black hole, we subtract the setup with $T=\tilde \delta=0$, such that
\be
f= \frac{F}{\int d^3 x} = I_T - I_0
\ee
where $I_T=S_{IIB}+S_{fl}+S_{GH}|_{\textrm{on-shell}}$ is the on-shell action (with the corresponding Gibbons-Hawking term) for the black hole solution and $I_0$ is the corresponding quantity in the case with no black hole but periodic Euclidean time. The periodicity of the time in the last case can be arbitrary, however,  we must ensure that the geometries we are comparing are matched at the UV cutoff (which we eventually send to infinity), and therefore the periodicity of the Euclidean time in the $\delta=0$ setup, $\beta_0$, is related to the periodicity of the black hole Euclidean time, $\beta$, as $\beta_0 = \sqrt{b(r_s)}\beta$. With this scheme the resulting free energy is finite\cite{D3D7QGP} and given by
\be
f=\varepsilon-Ts \ .
\ee
 Using the expressions \eqref{JT.gaugerunning} and \eqref{JT.dddTmu},  $-\partial f / \partial T=s$ holds, resulting in a non-trivial cross-check.\\

To study the grand-canonical ensemble we must consider the Legendre transform (with respect to $A_t'$) version of \eqref{1Dlagrangian}. This operation is accounted for by just a change of sign in the $F_{123}^2$ term, and a similar discussion to the one performed before for the canonical ensemble leads to
\be
\omega = \varepsilon - T s-\mu\, n_q \ ,
\ee
 and $-\partial \omega / \partial T=s$. In this case, the variation of $\tilde \delta$ with the temperature is determined by expression \eqref{JT.dddTnq}.\\

Let us summarize what we have attained so far. Given the  perturbative solution presented in \eqref{solgtt}--\eqref{solj} for the intersection of D3 and D7-branes in the Veneziano limit, we have succeeded in the calculation of the thermodynamic potentials in the canonical and grand-canonical ensembles by the use of the holographic dictionary. The explicit confirmation of the thermodynamic relations depends crucially on the variation of $\tilde \delta$ with the temperature, which is an ensemble dependent property, and on the running of the gauge coupling due to the non-vanishing beta function that can be deduced from \eqref{JT.gaugerunning}.\\

Once the  potentials in the (grand-)canonical ensembles are known we can proceed to study the thermodynamic stability of the solution. A first check consists on determining the determinant of the matrix of susceptibilities\\
\be
\Sigma = - \begin{pmatrix} \omega_{,T,T} & \omega_{,\mu,T} \\ \omega_{,T,\mu} & \omega_{,\mu,\mu}  \end{pmatrix} \ ,
\ee
where the comma stands for a partial derivative. This matrix is symmetric and $-\omega_{,\mu,\mu}\equiv \chi$ is the standard quark susceptibility. The components of the matrix are\\
\bea
-\omega_{,T,T} & = & \frac{3\pi^5}{2{\mathrm{Vol}}(X_5)}\ N_c^2\ T^2 \left[1+\frac12\epsilon_h + \frac{11}{24}\epsilon_h^2 \right] \nonumber\\
 &  & +\frac{\pi {\mathrm{Vol}}(X_3)}{4{\mathrm{Vol}}(X_5)}\ N_fN_c\ \mu^2\ \left[1+\frac{1}{6} \epsilon_h \right]  , \\
-\omega_{,\mu,T} & = & \frac{\pi {\mathrm{Vol}}(X_3)}{2{\mathrm{Vol}}(X_5)}\ N_fN_c\ \mu\ T \left[1+\frac{1}{6}\epsilon_h \right]  \ , \\
-\omega_{,\mu,\mu} & = & \frac{\pi{\mathrm{Vol}}(X_3)}{4{\mathrm{Vol}}(X_5)}\ N_fN_c\ T^2 \left[1+\frac{1}{6} \epsilon_h \right] \ ,\eea
and the determinant is parametrically positive, signaling thermodynamic stability.\\

In appendix \ref{appendix:susceptibility} it is shown that for ${\cal N}=4$ SYM (${\mathrm{Vol}}(X_3)=2\pi^2, {\mathrm{Vol}}(X_5)=\pi^3$) at zero coupling $\chi(\lambda=0)= N_fN_cT^2$, so that up to first order in $\epsilon_h$ we have $\chi(\lambda\rightarrow \infty)/\chi(\lambda=0)= \frac12 [1+\frac16 \epsilon_h]$.
This means that the effect of the dynamical flavors is to \emph{increase} the ratio of the susceptibilities of the interacting over free theory, pushing it above the value $1/2$.
The analogous situation in QCD is such that in the quenched case this ratio is close to one at relatively small temperatures above $T_c$\cite{Cheng:2009zi} (see also the comments in section 3.1.1 of Ref. \onlinecite{CasalderreySolana:2011us}), while in the unquenched case the ratio is \emph{reduced} to approximately $3/4$ at $T\sim 2T_c$.\cite{Borsanyi:2010bp}
Thus, on one side the effects of dynamical flavors is qualitatively opposite in the two cases (flavored ${\cal N}=4$ SYM and QCD); on the other side, they make the numerical values of the ratios in the two theories closer to each other, much like the corresponding ratios concerning other thermodynamic variables ($s, \varepsilon, \omega$). \\

We can study also the change of the internal energy as we vary the temperature, described by the heat capacity of the system, $c_V=\partial \varepsilon/\partial T$. The specific heat can be determined both at fixed chemical potential $\mu$ or quark density $n_q$ and, for a thermodynamically stable phase, the result should be positive. Indeed, a direct evaluation considering \eqref{JT.dddTnq} or \eqref{JT.dddTmu}, depending on whether we fix $n_q$ or $\mu$, gives the positive result
\bea
c_{V,n_q}  & = &  \frac{3\pi^5 N_c^2 T^3  }{2{\mathrm{Vol}}(X_5)}   \left[ 1+ \frac{\epsilon_h}{2}  \left( 1-\tilde \delta^2 \right) + \frac{\epsilon_h^2}{24} \left( 11-7\tilde \delta^2 \right)  \right]  \ , \nonumber \\
c_{V,\mu}  & = &  \frac{3\pi^5 N_c^2 T^3  }{2{\mathrm{Vol}}(X_5)}   \left[ 1+ \frac{\epsilon_h}{2}  \left( 1 +\tilde \delta^2 \right) + \frac{\epsilon_h^2}{24} \left( 11+7\tilde \delta^2 \right)  \right]  \ .\nonumber
\eea

With the heat capacity at constant chemical potential we can read off the speed of sound\\
\be\label{speedofsound}
c_s^2 = \frac{s}{c_{V,\mu}} = \frac{1}{3} \left( 1 - \frac{\epsilon_h^2}{6} \right) \ .
\ee
This result is parametrically positive and always lower than the conformal result $c_s^2=1/3$, in agreement with proposals in Ref. \onlinecite{cherman}. Consequently, we see that conformality is broken at order $\epsilon_h^2$ by the addition of fundamental matter. The fact that this effect is not seen at linear order in $\epsilon_h$ is due to considering massless  flavor degrees of freedom.  Furthermore, the deviation from conformality is  not affected by $\tilde \delta$. To confirm this point we can obtain the interaction energy of the system $\varepsilon-3p$ (where $p$ is the pressure), which is a quantity that vanishes for a conformal theory in three spatial directions. Using $p=-\omega$ we obtain
\be
\frac{\varepsilon-3p}{T^4} = \frac{\pi^2 N_c^2}{16 \mathrm{Vol}(X_5)}\epsilon_h^2 \ ,
\ee
which yet again is proportional to $\epsilon_h^2$ and independent of $\tilde \delta$.

\section{Hydrodynamics}
\label{sectionhydro}

In the D3-D7 field theories with massless flavors which are the subject of the present work, conformal invariance is broken at the quantum level. The breaking is due to marginally irrelevant terms in the action, accounting for the coupling of the fundamental flavor fields (introduced by means of the D7-branes) with adjoint (or bifundamental) fields of the quiver theory on the D3-branes. In the previous sections we have shown how to treat the conformality breaking effects in a perturbative expansion in the small parameter $\epsilon_h$. In the following, focusing on the simpler uncharged case, we will review how in such a scenario all the hydrodynamic transport coefficients, up to second order in the hydrodynamical derivative expansion,  can be explicitly evaluated.\cite{hydro1,hydro2} The coefficients are given in terms of a single parameter (e.g. the speed of sound) and the results extend to any holographic plasma where conformal invariance is slightly broken by a small marginally (ir)relevant deformation dual to a single scalar field in the bulk. To these systems, which we will call ``holographic marginal plasmas'' we devote the following section.
\subsection{Holographic Marginal Plasmas}
Let us consider a 4d strongly coupled plasma dual to a simple 5d gravity action including just the metric and a scalar field $\varphi$ with some potential $V(\varphi)$
\begin{equation}
S=\frac{1}{2\kappa_5^2}\int d^5x \sqrt{g}\left[R - \frac12(\partial\varphi)^2 - V(\varphi)\right]\,.
\end{equation}
It can happen that the potential has an $AdS_5$ (black hole) minimum at $\varphi=const.$ (say, at $\varphi=0$), i.e. that $V(\varphi)\approx -12 + (m^2/2)\varphi^2 + {\cal O}(\varphi^3)$ (setting the $AdS$ radius to one). In this case we know that the scalar field is holographically dual to a 4d operator whose dimension $\Delta$, around the conformal fixed point dual to the $AdS$ background, is given by $\Delta(\Delta-4)=m^2$. If $\Delta<4$ (resp. $\Delta>4$), the operator is relevant (resp. irrelevant). If $\Delta=4$ the operator is exactly marginal and does not drive the theory away from the conformal point. Examples of exactly marginal deformations are the ``$\beta$-deformations''.\cite{Lunin:2005jy}\\

Marginally (ir)relevant operators, instead, break conformal invariance, driving logarithmic flows of the couplings, despite their classical dimension being equal to four. The way to holographically account for such operators in the simplified model considered here is the following: let us assume that the potential, for small values of some parameter $\gamma$, has a leading linear term
\be
V(\varphi)\approx -12(1 + \gamma\varphi) + {\cal O}(\gamma^2\varphi^2)\,.
\ee
At leading order the scalar field is thus massless and corresponds to an operator of classical dimension four. Moreover the 5d action does not admit an $AdS$ minimum with $\varphi=0$, if $\gamma\neq0$. If $|\gamma|\ll1$ we can easily find a perturbative solution for metric and scalar. At zeroth order in $\gamma$ the metric is $AdS$ (black hole) and, say, $\varphi=0$. At first order, the equation of motion for the scalar field is solved (assuming $\varphi$ to depend only on the radial variable of the background) by
\be
\varphi(r) = -3\gamma\log\frac{r}{r_h}\,.
\ee
Here we have considered the non-extremal background with horizon radius $r_h$.\\

The logarithmic running of the scalar field and the $AdS$/CFT radius/energy relation $r=E$ (which is still valid at leading order in $\gamma$) essentially realize the logarithmic RG flows expected in the 4d theory. If, as it happens in the simplest models where $\varphi$ is the dilaton, the 4d running coupling is holographically given by $g^2_{4d}=e^{\varphi}$, then models with $\gamma>0$ (resp. $\gamma<0$) are in the marginally relevant (resp. irrelevant) class.\\

There are many known string embeddings of the simple model considered above, extending from examples with ``fractional'' branes\cite{klebnek} in type 0B string theory\cite{ktzero} and in the type IIB conifold model\cite{ks} - which sit in the marginally relevant class -, to the examples with flavor D7-branes considered in the present review, which are in the marginally irrelevant class. In these cases, with just one parameter driving the deformation (roughly speaking, the number of fractional or flavor D-branes added to an otherwise conformal model), the 5d gravity action effectively reduces to the single scalar one considered above when the parameter is taken to be very small. This scalar is a string modulus dual to the marginally (ir)relevant term in the 4d field theory action. In general, of course, the 5d reduction of a type IIB gravity action gives rise to various scalar fields, but at leading order it is just this scalar which plays an active role. The other ones, at
 leading order, are just ``frozen'' at their zeroth order values. For the D3-D7 uncharged plasma\cite{D3D7QGP} and the cascading conifold one,\cite{Gubser:2001ri} this has been explicitly shown in Ref. \onlinecite{hydro1}. We report the analysis for the uncharged D3-D7 plasma in appendix \ref{appendixB}, from which it follows that, at leading order,
\be
\varphi=\Phi-\Phi_h\,, \qquad \gamma=-\frac{\epsilon_h}{3}\,,
\ee
where $\Phi$ is the dilaton.
\subsection{Equilibrium Properties: Equation of State}
One of the simplest way to realize a scalar potential going linearly at leading order is by choosing it to have an exponential form, $V(\varphi)=-12e^{\gamma\varphi}$, as it happens in the Chamblin-Reall models.\cite{CR} Of course, the details of the non-leading terms in $\gamma$ are irrelevant if one wants to focus on a leading order analysis. Anyway, the Chamblin-Reall models have a first advantage of being exactly solvable and so, in particular, an exact black-hole solution is known to all orders in $\gamma$. The thermodynamics of the dual 4d models thus easily follows. In particular, as it has been discussed in Ref. \onlinecite{gubserspeed}, the speed of sound is given by\\
\be
c_s^2 = \frac{dp}{d\varepsilon}= \frac13 -\frac{1}{2} \frac{V'(\varphi)^2}{V(\varphi_h)^2} = \frac13 - \frac{\gamma^2}{2}\,,
\label{velsuono}
\ee
so that the simple equation of state\\
\be
p = \left(\frac13 - \frac{\gamma^2}{2}\right)\varepsilon\,,
\ee
follows; here $p$ is the pressure and $\varepsilon$ is the energy density. These expressions explicitly show how the parameter $\gamma$ weighs the conformality breaking effects: for $\gamma=0$ we get back the conformal result $c_s^2=1/3$, i.e. $\varepsilon=3p$.
Indeed, using $\gamma=-\epsilon_h/3$, we see that the speed of sound precisely matches that obtained for D3-D7 plasmas with $\epsilon_h\ll1$ as discussed in the previous sections.\\

As has been discussed in Refs. \onlinecite{buchelbound,Kanitscheider:2009as}, Chamblin-Reall models in $d+1$ dimensions have a further nice property: for particular values of the coefficient of the exponential in the potential, they can be obtained from dimensional reduction on a $2\sigma-d$ torus of Einstein actions with negative cosmological constant in $2\sigma+1$ dimensions. This happens when the parameter $\sigma$, which determines the coefficient in the exponential together with $d$, is half-integer. For these values of $\sigma$, one can then start from the well-known $AdS_{2\sigma+1}$ solution and its dual energy-momentum tensor (at or near equilibrium) and obtain the energy-momentum tensor for the dual to the Chamblin-Reall model by simple toroidal dimensional reduction.\\

For instance, let us just focus on the equilibrium properties in the $d=4$ case. The higher dimensional action has an $AdS_{2\sigma+1}$ (black hole) minimum, dual to a $2\sigma$-dimensional (thermal) CFT. The equation of state for the latter is just given by $\varepsilon=(2\sigma-1)p$ as it follows from the vanishing trace of the stress-energy tensor. The toroidal dimensional reduction does not affect the equation of state which will thus be the same for the 4d plasma dual to the Chamblin-Reall model. Using the previous result, we see that $2\sigma-1=6/(2-3\gamma^2)$. Thus, if $\gamma\ll1$, $\sigma\approx 2+ (9/4)\gamma^2$. As we have previously recalled, the analysis requires $\sigma$ to be a semi-integer and this is certainly not realized for any $\gamma$.\\

However, a crucial observation in Ref. \onlinecite{Kanitscheider:2009as} is that, from the point of view of the theory in $d+1$ dimensions, the equations are smooth in the parameter $\sigma>d/2$ (at $\sigma=d/2$ the action is singular\cite{Kanitscheider:2009as}). Then one can proceed as follows. One starts from a Chamblin-Reall model in $d+1$ dimensions for whatever $\sigma>d/2$ and performs the continuation (which is smooth) to the nearest value $\tilde\sigma$ which is semi-integer. The latter theory is the compactification of a theory admitting a $AdS_{2\tilde\sigma+1}$ solution, so its dual energy-momentum tensor, which will be a function of $\tilde\sigma$, can be calculated straightforwardly. This  energy-momentum tensor can thus be continued (smoothly) back to the one of the theory corresponding to the original value $\sigma$. For the case $d=4$ and $\gamma\ll1$ we can thus choose $\tilde\sigma=5/2$.\\

Notice that this procedure allows to get the full stress energy tensor at or near equilibrium. In the latter case, we can thus deduce the full transport coefficients of the 4d plasma dual to the 5d Chamblin-Reall model, just by dimensional reduction of the hydrodynamic stress tensor of an holographic higher dimensional conformal plasma.
\subsection{Hydrodynamics: Generalities}
Hydrodynamics is the effective theory of long wavelength, low frequency fluctuations around local thermodynamic equilibrium. In the uncharged case relativistic hydrodynamics is determined, up to second order in the derivative expansion, by seventeen transport coefficients, fifteen of which are possibly independent.\cite{Baier:2007ix,Bhattacharyya:2008jc,Romatschke:2009kr}
On a general space with metric $g_{\mu\nu}$, the energy momentum tensor
\begin{equation}\label{tmunu}
T^{\mu\nu}=\varepsilon u^\mu u^\nu + p  \Delta^{\mu\nu} + \pi^{\mu\nu} + \Delta^{\mu\nu}\Pi\,,
\end{equation}
where $\Delta^{\mu\nu}=g^{\mu\nu}+u^\mu u^\nu$, is determined by the energy density $\varepsilon$, fluid velocity $u^\mu$ ($u^\mu u_\mu=-1$), the transport coefficients in its ``viscous shear'' part:
\begin{eqnarray}\label{shear}
&&\pi^{\mu\nu}= -\eta \sigma^{\mu\nu} +\eta \tau_\pi \Bigl[\langle D \sigma^{\mu\nu}\rangle + \frac{\nabla \cdot u}{3}\sigma^{\mu\nu} \Bigr] + \nonumber \\
&&+\kappa \Bigl[ R^{<\mu\nu>}-2 u_\alpha u_\beta R^{\alpha <\mu\nu> \beta} \Bigr] + \lambda_1 \sigma^{<\mu}_{\lambda} \sigma^{\nu>\lambda} +  \nonumber \\
&&+ \lambda_2 \sigma^{<\mu}_{\lambda} \Omega^{\nu>\lambda} + \lambda_3 \Omega^{<\mu}_{\quad \lambda} \Omega^{\nu>\lambda}  + \kappa^* 2 u_\alpha u_\beta R^{\alpha <\mu\nu> \beta} +  \nonumber \\
&&+ \eta \tau_\pi^* \frac{\nabla \cdot u}{3}\sigma^{\mu\nu} + \lambda_4 \nabla^{<\mu} \log{s}  \nabla^{\nu >} \log{s}\, ,
\end{eqnarray}
and in its ``viscous bulk'' part:
\begin{eqnarray}\label{bulk}
&&\Pi =-\zeta (\nabla \cdot u) + \zeta \tau_\Pi D  (\nabla \cdot u) + \xi_1  \sigma^{\mu\nu}\sigma_{\mu\nu}+ \xi_2  (\nabla \cdot u)^2 + \nonumber \\
&&+\xi_3 \Omega^{\mu\nu}\Omega_{\mu\nu} + \xi_4 \nabla_{\mu}^{\perp} \log{s} \nabla^{\mu}_{\perp} \log{s}+ \nonumber \\
&& + \xi_5 R + \xi_6 u^\alpha u^\beta R_{\alpha \beta}\,,
\end{eqnarray}
while the pressure is given by the equation of state $p(\varepsilon)$.
The various structures in (\ref{shear}) and (\ref{bulk}), apart from the obvious Riemann and Ricci tensors and scalar curvature ($R^{\mu\nu\rho\sigma}, R^{\mu\nu}, R$), are given by
\begin{eqnarray}
D &\equiv & u^\mu\nabla_\mu\,, \quad \nabla^{\mu}_{\perp} \equiv  \Delta^{\mu\nu}  \nabla_{\nu}\,,\nonumber\\
\sigma^{\mu\nu}&\equiv & \nabla^{\mu}_{\perp} u^\nu + \nabla^{\nu}_{\perp} u^\mu -\frac23  \Delta^{\mu\nu}(\nabla \cdot u)\,, \nonumber\\
\Omega^{\mu\nu} &\equiv & \frac12 (\nabla^{\mu}_{\perp} u^\nu - \nabla^{\nu}_{\perp} u^\mu)\,, \qquad
\end{eqnarray}
and for a generic tensor $A^{\mu\nu}$ the notation was used that
\begin{equation}
\langle A^{\mu\nu} \rangle = A^{<\mu\nu>} \equiv \frac12  \Delta^{\mu\alpha}\Delta^{\nu\beta}(A_{\alpha\beta}+A_{\beta\alpha})-\frac13  \Delta^{\mu\nu} \Delta^{\alpha\beta}A_{\alpha\beta}\,.
\end{equation}
Finally, $s$ is the entropy density, while $c_s^2=dp/d\varepsilon$ is the square of the speed of sound.\\

The shear viscosity $\eta$ and the second order coefficients $\tau_\pi$ (``shear'' relaxation time), $\kappa$, $\lambda_1, \lambda_2, \lambda_3$ are the only ones defined in conformal fluids, such as the one of ${\cal N}=4$ SYM.
All the others coefficients, i.e. the bulk viscosity $\zeta$ and the second order coefficients $\kappa^*, \tau_\pi^*, \lambda_4, \tau_\Pi$ (``bulk'' relaxation time), $\xi_1,   \xi_2,  \xi_3,  \xi_4,  \xi_5,  \xi_6$, are only defined in non-conformal plasmas.
\subsection{Hydrodynamics of holographic marginal plasmas}\label{section:coefficients}
Just like any theory with a dual gravity description satisfying the general requirements established in Ref. \onlinecite{Kovtun:2004de}, 4d marginal plasmas dual to (small $\gamma$) 5d Chamblin-Reall models have a shear viscosity over entropy density ratio given by (in units $\hbar=K_B=1$)
\begin{equation}
\frac{\eta}{s}=\frac{1}{4\pi}\,.
\end{equation}
Moreover, according to the results in Ref. \onlinecite{gubserspeed}, these models have a bulk viscosity given by
\be
\frac{\zeta}{\eta}= \left(\frac{V'(\varphi)}{V(\varphi)}\right)^2 = \gamma^2\,,
\label{zetagub}
\ee
so that in the D3-D7 case, $\frac{\zeta}{\eta}=\epsilon_h^2/9$ at leading order (in appendix \ref{appendixB} we will cross-check this result using a general holographic formula for the bulk viscosity proposed in Ref. \onlinecite{Eling:2011ms}). Comparing this expression with that of the speed of sound in eq. (\ref{velsuono}) we see that the bulk viscosity saturates the bound
\be
\frac{\zeta}{\eta}\ge 2 \left(\frac{1}{d-1} - c_s^2\right)\,,
\ee
proposed in Ref. \onlinecite{buchelbound} for any $d+1$ dimensional plasma with gravity dual. As observed in Refs. \onlinecite{hydro1,hydro2} this is a generic feature of holographic marginal plasmas at leading order.\\

In order to compute the second order hydrodynamic transport coefficients, one can take advantage from the fact that, as previously discussed, for $\gamma\ll1$, 5d Chamblin-Reall solutions can be obtained from reduction and analytic continuation (from $\tilde\sigma=5/2$ to $\sigma\approx 2+ (9/4)\gamma^2$) of an $AdS_{2\tilde\sigma+1}$ background, whose dual conformal hydrodynamics was considered in Ref. \onlinecite{Bhattacharyya:2008mz}. Let us skip the details of the (simple) derivation, for which we refer to the original paper\cite{hydro2} (see also Refs. \onlinecite{springer,Romatschke:2009kr}).\\

The final result is shown in table \ref{relations}, where we give the full list of transport coefficients for a holographic marginal plasma, at leading order in the conformality breaking parameter
\begin{equation}\label{Delta}
\delta_{cb}\equiv \frac32\gamma^2 = (1-3c_s^2)\,.
\end{equation}
In particular to get the transport coefficients of the uncharged flavored D3-D7 plasmas, to second order in $\epsilon_h$, it suffices to notice that
\be
\delta_{cb}(D3-D7) = \frac{\epsilon_h^2}{6}\,.
\ee
According to what has been previously commented (see also Ref. \onlinecite{hydro1}), the results in table \ref{relations} automatically apply to the cascading plasma as well,\cite{Gubser:2001ri} at leading order in $\delta_{cb}$.\\
\begin{widetext}
\begin{center}
\begin{table}[h]
\begin{tabular}{||c|c||c|c||c|c||}
\hline
 & & & & & \\
$ \frac{\eta}{s} $ & $\frac{1}{4\pi}$ &  $T\tau_{\pi}  $  & $ \frac{2-\log{2}}{2\pi} + \frac{3(16-\pi^2)}{64\pi}\delta_{cb} $  & $ \frac{T\kappa}{s} $  &  $  \frac{1}{4\pi^2}\Bigl(1-\frac34 \delta_{cb} \Bigr) $  \\
 & & & & & \\
\hline \hline
 & & & & & \\
$\frac{T \lambda_1}{s}  $ & $\frac{1}{8\pi^2}\Bigl(1+\frac34 \delta_{cb} \Bigr) $ & $\frac{T \lambda_2}{s} $ & $-\frac{1}{4\pi^2}\Bigl( \log{2}+\frac{3\pi^2}{32}\delta_{cb} \Bigr) $ & $\frac{T \lambda_3}{s} $ & $0 $ \\
 & & & & & \\
\hline \hline
 & & & & & \\
$\frac{T\kappa^*}{s} $ & $-\frac{3}{8\pi^2}\delta_{cb} $ & $T\tau_{\pi}^* $ & $-\frac{2-\log{2}}{2\pi}\delta_{cb} $ & $\frac{T \lambda_4}{s}  $ & $0 $ \\
 & & & & & \\
\hline \hline
 & & & & & \\
$\frac{\zeta}{\eta} $ & $\frac23 \delta_{cb} $ & $T\tau_{\Pi} $ & $\frac{2-\log{2}}{2\pi} $ & $\frac{T \xi_{1}}{s} $ & $\frac{1}{24\pi^2}\delta_{cb} $ \\
 & & & & & \\
\hline \hline
 & & & & & \\
$ \frac{T \xi_{2}}{s} $ & $\frac{2-\log{2}}{36\pi^2}\delta_{cb} $ & $\frac{T \xi_{3}}{s} $ & $0 $ & $\frac{T \xi_{4}}{s} $ & $0 $ \\
 & & & & & \\
\hline \hline
 & & & & & \\
$\frac{T \xi_{5}}{s} $ & $\frac{1}{12\pi^2}\delta_{cb} $ & $\frac{T \xi_{6}}{s} $ & $\frac{1}{4\pi^2}\delta_{cb} $ & & \\
 & & & & & \\
\hline
\end{tabular}
\caption{The transport coefficients, in the notation of (\ref{tmunu})-(\ref{bulk}), for a marginally (ir)relevant deformation of a conformal theory, at leading order in the deformation parameter $\delta_{cb} \equiv (1-3c_s^2)$. The holographic equation of state is $\varepsilon=3(1+\delta_{cb})p$. For D3-D7 plasmas $\delta_{cb}=\epsilon_h^2/6$.}\label{relations}
\end{table}
\end{center}
\end{widetext}
A comment has to be devoted to the two relaxation times $\tau_{\pi}, \tau_{\Pi}$. At leading order in the conformality breaking, the bulk relaxation time $\tau_{\Pi}$ is not proportional to the bulk viscosity. The shear relaxation time $\tau_{\pi}$, moreover, depends on the speed of sound. Interestingly, the magnitude of the shear relaxation time $\tau_{\pi}$ is larger than the one of the bulk relaxation time $\tau_{\Pi}$; moreover, if we naively extrapolate the qualitative temperature dependence of the QCD speed of sound to the present formulas, $\tau_{\pi}$ is found to increase more steeply than $\tau_{\Pi}$ as the temperature is reduced, contrary to the common speculations about the behavior of these two coefficients.
The discrepancy could depend on the strong coupling regime we are considering.\\

In addition, using the above results it is easy to verify that the relation
\begin{equation}
4\,\lambda_1 + \lambda_2 = 2\,\eta\, \tau_{\pi}\,,
\label{relhy}
\end{equation}
holds, at first order in $\delta_{cb}$. It has been shown in Refs. \onlinecite{erd,hy} that (\ref{relhy}) is satisfied in all the known examples of conformal plasmas (in $d\ge 4$ spacetime dimensions, with or without conserved global charges) with a dual gravity description. Our results provide a unique validity check of the above relation in non-conformal settings.\\

As an independent check of the results shown in table \ref{relations}, we will provide, in appendix \ref{alternative}, a direct holographic computation of some (combinations of) the first and second order transport coefficients.\\

Let us conclude by mentioning that in Ref. \onlinecite{hydro2} holographic marginal plasmas have been used as toy models for the initial stages of the evolution of the Quark-Gluon Plasma produced in heavy ion collisions at RHIC and LHC. Indeed, if the temperature of the plasma is somewhat higher than the temperature for the QCD crossover between confined and deconfined phases, it is sensible to approximate the QGP as a slightly deformed conformal fluid. The results in table \ref{relations} thus give all the transport coefficients in the toy model, once only the speed of sound at some temperature is given (from lattice data).

\subsection{Negative Refraction and Additional Light Waves}\label{optics}

Thermodynamic quantities and hydrodynamic transport coefficients are sufficient to determine quantitatively some interesting optical properties of the D3-D7 plasmas.
With ``optical properties'' we mean the response of the system to an external $U(1)_{em}$ electromagnetic incident wave.
At leading order in the electromagnetic coupling $q$, the $U(1)_{em}$ can be treated as global and we can use our charged solution as a model of an electrically charged, strongly coupled relativistic plasma, i.e. we treat $U(1)_B$ as the $U(1)_{em}$, and the fundamental matter as $U(1)_{em}$-charged particles.
With respect to the R-symmetry charged solutions, the model at hand has the advantage that, as in ordinary ``phenomenological'' systems, only a fraction of the fluid constituents is charged.\\

As a first optical property, it has been shown in Refs. \onlinecite{ragazzi1,Amariti:2011dj} that any charged plasma exhibits negative refraction of light for small enough frequencies (holographic studies also appear in Refs. \onlinecite{Gao:2010ie,Ge:2010yc,Bigazzi:2011it,Bigazzi:2011ut,Amariti:2011dm}).
This means that the phase velocity and the energy flux have opposite directions, resulting in a number of non-standard optical properties.
The critical frequency for the negative refraction is (we consider here only the ${\cal N}=4$ SYM case)
\begin{equation}\label{negref}
\omega^2_{crit} = 4\pi q^2 \frac{n_q^2}{\varepsilon+p} \sim \frac{8 q^2}{\pi^2 N_c^2}\frac{n_q^2}{T^4}\left[1-\frac12 \epsilon_h  \right]\,.
\end{equation}
Since $n_q/T^3 \sim {\cal O}(N_f N_c \sqrt{\lambda_h} \tilde \delta)$, the window of frequencies where negative refraction is present is ${\cal O}(N_f^2 \lambda_h \tilde\delta^2)$.\\

Moreover, it is known that the same phenomenon that gives rise to negative refraction, i.e. spatial dispersion, is responsible for the possible presence of so-called ``Additional Light Waves'' (ALW) in certain materials: for each incident light wave, there could be more than one propagating waves, differing for their refractive index.
This phenomenon has been studied holographically in Ref. \onlinecite{ragazzi2} in the R-symmetry charged solutions.
Again, in the plasma regime the phenomenon is (obviously) entirely determined by thermodynamics and transport coefficients.\\

\begin{figure}
\includegraphics{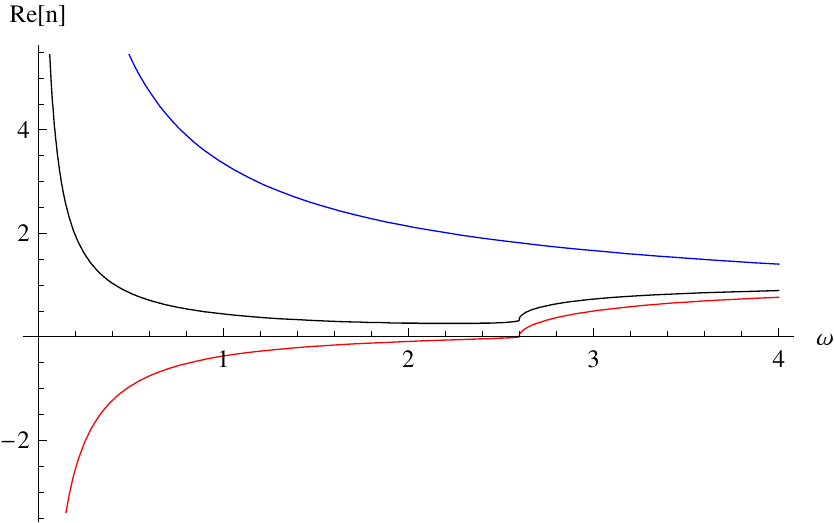}
\caption{\label{alw} Real part of the refractive index, as a function of the frequency, for the ALW (top line) and the normal wave (bottom line); the middle line is the ``effective refractive index''. In this plot we used the parameters $N_c=100, N_f=20, \lambda_h=50, q=1, \tilde\delta=0.1$.}
\end{figure}

For the solution at hand, the resulting effect is very similar to the R-symmetry charged case: using formulas in  Ref. \onlinecite{ragazzi2}, one can check the presence of at least one ALW besides the normal wave.
In figure \ref{alw} we report the plot of the real part of the refractive index Re[n] as a function of the frequency $\omega$ for the two waves.
The bottom line in the plot refers to the ordinary wave, which exhibits negative refraction (signaled here by a negative Re[n]) for frequencies smaller than the critical one given in (\ref{negref}).
The upper line refers to the ALW, with positive refraction.
The middle line is an ``effective refractive index'' measuring the relative magnitude of the two waves: its interpolating behavior between the two other lines signals the fact that in the intermediate frequency regime both waves are propagating with comparable amplitudes.
Unfortunately, as in all the holographic models analyzed so far, the imaginary part of the refractive index is very large in the interesting frequency regime, producing a big dumping of the propagating waves.


\begin{acknowledgments}
We would like to express our gratitude to  Paolo Benincasa, Roberto Emparan, Veselin Filev, Gianluca Grignani, Gabriele Martelloni, Ioannis Papadimitriou and  S. Prem Kumar for very illuminating discussions.
The research of A.L.C. and F.B. is supported by the European Community Seventh Framework Programme FP7/2007-2013, under grant agreements n. 253534 and 253937. J.T. and D.M. are supported
by the Netherlands Organization for Scientic Research (NWO) under the FOM Foundation research program.
J.M. is supported by the MICINN and  FEDER (grant
FPA2008-01838), the Spanish Consolider-Ingenio 2010 Programme CPAN
(CSD2007-00042), and the Xunta de Galicia (Conselleria de Educacion
and grant INCITE09-206-121-PR). Part of their research was done while J.M. and J.T. were visiting
the  Kavli Institute Beijing. They want to thank  KITPC Beijing
for hospitality and financial support (this last under grant KJCX2.YW.W10 of the Chinese
Academy of Sciences).

{ \it F. B. and A. L. C. would like to thank the Italian students,
parents, teachers and scientists for their activity in support of
public education and research.}
\end{acknowledgments}

\appendix
\section{The $U(1)_B$ Susceptibility of the Free Flavored ${\cal N}=4$ SYM Plasma}\label{appendix:susceptibility}

Let us consider the ${\cal N}=2$ theory obtained from ${\cal N}=4$ $SU(N_c)$ SYM coupled with $N_f$ massless fundamental hypermultiplets. The fields in the latter have the same charge $q=1$ under the baryonic $U(1)_B$. All the other adjoint fields in the theory are naturally uncharged.\\

In order to compute the total $U(1)_B$ susceptibility\cite{mart}
\be
\chi = \frac{\partial n_B}{\partial \mu}\,({\rm at}\,\mu=0)\,,
\label{sus}
\ee
for the free theory, we just need the expression for the net total baryon charge density $n_B$. For each field the net charge density is given as the difference between the particle and the antiparticle density. Thus, for complex bosons of charge $q$ under a $U(1)$
\be
n_b = q \int\frac{d^3k}{(2\pi)^3}\left[F_b(E_k- q \mu) - F_b(E_k+ q \mu)\right]\,,
\ee
and for Weyl fermions (note: for Dirac fermions there is a further overall factor of two)
\be
n_{f} = q \int\frac{d^3k}{(2\pi)^3}\left[F_f(E_k- q \mu) - F_f(E_k+ q \mu)\right]\,.
\ee
In the expressions above $E_k = |k|\equiv k$ since we want to consider massless particles. Moreover
\be
F_{b}(x)=\frac{1}{e^{\beta x} -1}\,,\qquad F_{f}= \frac{1}{e^{\beta x}+1}\,,
\ee
are the standard Bose and Fermi distributions with $\beta=1/T$.\\

From the expressions above it is easy to compute the susceptibilities using the definition (\ref{sus}). The results are
\be
\chi_b = q^2 \frac{T^2}{3}\,,\qquad \chi_f = q^2 \frac{T^2}{6}\,.
\ee
In our case, counting the fields charged under $U(1)_B$, we have $2N_cN_f$ complex scalars and $2N_cN_f$ Weyl fermions. Thus the free baryon number susceptibility is ($q=1$ here)
\be
\chi(\lambda=0) = 2 N_c N_f \left(\frac{1}{3}+\frac{1}{6}\right)T^2= N_c N_f T^2\,.
\ee
In the planar infinite 't Hooft coupling limit the same theory, in the quenched approximation, has (see Ref. \onlinecite{myers})
\be
\chi(\lambda\rightarrow\infty)= \frac{1}{2} N_c N_f T^2 + \dots \,,
\ee
so that $\chi(\lambda\rightarrow\infty)/\chi(\lambda=0)=1/2+ \dots$ as it happens for the R-charged case (see for example appendix A in Ref. \onlinecite{teaney} and the comments in Ref. \onlinecite{CasalderreySolana:2011us}).\\

Just for comparison, let us consider the free  $SU(N_c)$ QCD with $N_f$ fundamental Dirac fermions. The total $U(1)_B$ susceptibility of the related Stefan-Boltzmann quark-gluon gas is readily evaluated as
\be
\chi_{QCD}(SB) = N_c N_f\, 2\, \frac{T^2}{6} = \frac{1}{3}N_c N_f T^2\,.
\ee

\section{The 5d Effective Action in the Uncharged Case}\label{appendixB}

The ten dimensional action (\ref{setup:theaction}), (\ref{setup:flavoraction}) can be reduced to a five dimensional effective action with a suitable ansatz for the various fields.
We will only discuss the uncharged case, so that ${\cal F}=F_{(3)}=B=0$.
Moreover, $F_{(1)}=Q_f(d\tau+A_{KE})$ and $F_{(5)}=Q_c(1+*)\epsilon_5$ (the last symbol is of course the volume form of the internal manifold).
The reduction ansatz of the metric is of the form\cite{Benini:2006hh}
\begin{eqnarray}\label{ansatz}
ds_{10}^2 &=&  e^{\frac{10}{3}f}g_{\mu\nu} dx^\mu dx^\nu + e^{-2(f+w)}ds_{KE}^2 \nonumber \\
&& + e^{2(4w-f)} (d\tau+A_{KE})^2\,.
\end{eqnarray}
We will use Greek indices $\mu, \nu$ for the five dimensional space.
In the parametrization above
\begin{equation}
f = -\frac15\log\left( h^{\frac54} S^4 F \right)\, , \qquad w = \frac15 \log\left( \frac{F}{S} \right)\,  ,
\label{BCdef}
\end{equation}
and
\begin{eqnarray}
ds_{5}^2 & =& (h^{\frac{5}{4}} S^4 F)^{\frac{2}{3}} \left[-h^{-1/2}b\,dt^2 +h^{-1/2} d\vec x_3^2 \right. \nonumber \\
&& \left. + h^{1/2} S^8 F^2 b\,d\sigma^2 \right] \, .
\end{eqnarray}

Plugging the above ansatze in the ten dimensional action and integrating in the five internal direction, we obtain the five dimensional effective action\cite{Benini:2006hh}
\begin{eqnarray}
S_5 &=& \frac{{\mathrm{Vol}}(X_5)}{2\kappa_{10}^2} \int d^5 x \sqrt{-\det g} \left[ R[g] - \frac{40}{3}(\partial f)^2 - 20 (\partial w)^2 \nonumber \right. \\
&&\left. - \frac{1}{2}(\partial\Phi)^2 - V(\Phi,f,w)\right]\, ,
\label{5daction}
\end{eqnarray}
where
\begin{eqnarray}
V(\Phi,f,w) &=& 4 e^{\frac{16}{3}f+2w} \left( e^{10w}-6 + Q_f e^{\Phi}\right) + \nonumber \\
&&\frac{1}{2} Q_f^2 e^{\frac{16}{3} f -8w +2\Phi} + \frac{Q_c^2}{2}e^{\frac{40}{3} f}\, .
\label{potV2}
\end{eqnarray}
We have set $R_{AdS}=1$ and $\alpha'=1$ for simplicity. In these units $Q_c=4$.\\

Let us study the above 5d action perturbatively in $\epsilon_h$, writing each of the scalar fields as well as the 5d metric components as $\Psi=\Psi_0 + \epsilon_h \Psi_1 + ...$. At order zero the action has an $AdS$ (black hole) minimum at $f=w=0$ and $\Phi=const.\equiv \Phi_h$. The field $f$ is dual to an irrelevant operator of dimension $\Delta=8$ and the field $w$ is dual to a vev for an irrelevant operator of dimension $\Delta=6$. The dilaton $\Phi$ is dual to the insertion of a marginally irrelevant operator, actually the flavor term in the field theory ($T=0$) superpotential.\cite{Benini:2006hh} This is the source term which is responsible for the breaking of conformal invariance at the quantum level.\\

To better understand the role played by the various scalars, let us consider the quantity
\be
{\cal V}_\phi\equiv\left(\frac{V'(\phi_h)}{V(\phi_h)}\right)^2\,,
\label{calV}
\ee
in three different cases, where we identify $\phi$ with one of the three fields entering the potential (\ref{potV2}), taking the other two scalars fixed to their background values. At order $\epsilon_h^2$ we get
\begin{equation}
{\cal V}_{f,w}= 0 \, , \qquad {\cal V}_{\Phi}=\frac{\epsilon_h^2}{9}\, .
\end{equation}
This indicates\cite{hydro1} that, among the three scalar fields in the action, only the dilaton - i.e precisely the field dual to the source for the (marginally irrelevant) deformation driving our theories away from conformality - plays the role of ``active'' field in the game at leading order. The other two scalars, $f$ and $w$, do not contribute at leading order and they can be fixed to their background values.\\

All in all, at leading order in the conformality breaking parameter, the system is effectively captured by a simple 5d dual gravity model with a single scalar field with linear potential. Actually, since $\epsilon_h^2$ is a constant up to higher order terms, the effective 5d action, at leading order, is mapped (see section \ref{sectionhydro}) into a Chamblin-Reall model\cite{CR} with constant $\gamma=-\frac{\epsilon_h}{3}$ (the sign is chosen in order for the solution to the scalar field equation to match the one found for the dilaton in the main body of the paper).\\

Using the holographic field/operator map found in Ref. \onlinecite{bgz}, in Ref. \onlinecite{hydro1} it was shown that analogous considerations hold for the cascading plasma too, at leading order in the conformality breaking parameter.
\subsection{The Bulk Viscosity From a General Holographic Formula}
As a further check of the previous conclusions, which in section \ref{sectionhydro} have been heavily used to study the hydrodynamics of the D3-D7 plasmas, let us consider a general holographic formula for the bulk viscosity
\begin{equation}
\frac{\zeta}{\eta}=\sum_i\left(s\frac{d\phi^h_i}{ds} + \rho^a\frac{d\phi^h_i}{d\rho^a}\right)^2\,,
\label{oz}
\end{equation}
found in Refs. \onlinecite{Eling:2011ms,Eling:2011ct}. In this expression the suffix $h$ means ``evaluated at the horizon'', $\phi_i$ are the scalar fields with canonically normalized kinetic term in the 5d gravity action, and $\rho^a$ is the charge density holographically related to the $a-th$ U(1) field in the 5d action. In our case we have just the $U(1)_B$ charge. Finally, $s$ is the entropy density.\\

Let us just consider the uncharged case for simplicity. The scalar fields in the 5d action are $\Phi,f,w$. Their background values are $\Phi\approx\Phi_h +\epsilon_h \log(r/r_h)$, $f\approx -\epsilon_h/40$, $w\approx -\epsilon_h/60$ at leading order. Now $s d\Phi/ds\approx (r_h/3) (d\Phi/dr_h)\approx \epsilon_h/3$ up to subleading terms. The analogous quantity for $f$ and $w$ is of order $\epsilon_h^2$ . From this we readily find that only the dilaton contributes to the bulk viscosity, and using eq. (\ref{oz}) we find $\zeta/\eta=\epsilon_h^2/9$, confirming the result found in section \ref{sectionhydro}.


\section{Alternative Computation of Transport Coefficients}
\label{alternative}

In this appendix we present the direct computation of some (combinations of) first and second order transport coefficients, providing an independent check of the results in section \ref{sectionhydro}.\cite{hydro1,hydro2}
Specifically, the pressure $p$, the transport coefficients $\eta$, $\kappa$ and a combination of the ``shear'' relaxation time $\tau_\pi$ and $\kappa^*$ are derived from the retarded correlator of the tensorial fluctuation of the fluid\cite{Baier:2007ix,Romatschke:2009kr}
\begin{equation}\label{retcorr}
G_R^{xy,xy}=p-i \eta \omega + \Bigl( \eta \tau_\pi -\frac{\kappa}{2} +\kappa^* \Bigr)\omega^2 -\frac{\kappa}{2}q^2 + {\cal O}(q^3,\omega^3)\, .
\end{equation}
In this expression $\omega$ is the frequency of the fluctuation and $q$ its momentum.
The shear viscosity $\eta$ can be extracted also from the dispersion relation of the vectorial (hydrodynamic) fluctuation mode
\be\label{vectorialdispersion}
\omega = -i\,\frac{\eta}{sT} q^2 + {\cal O}(q^3)\, .
\ee
Moreover, the speed of sound $c_s$, the bulk viscosity $\zeta$ and a combination of the ``shear'' and ``bulk'' relaxation times $\tau_\pi, \tau_\Pi$ are derived from the dispersion relation of the scalar (hydrodynamic) mode (sound channel)
\begin{equation}\label{vecdiff2}
\omega = c_s q - i \Gamma q^2 + \frac{\Gamma}{c_s}\Bigl(c_s^2\tau^{eff}-\frac{\Gamma}{2}\Bigr)q^3 + {\cal O}(q^4)\,,
\end{equation}
where
\begin{equation}
\Gamma=\frac{\eta}{sT} \left( \frac{2}{3} + \frac{\zeta}{2\eta} \right)\,,\qquad \tau^{eff}=\frac{\tau_{\pi}+\frac{3\zeta}{4\eta}\tau_{\Pi}}{1+\frac{3\zeta}{4\eta}}\, .
\end{equation}

The fluid modes are dual to fluctuations of the five dimensional gravity fields above the solution of section \ref{sec:solution} in the zero charge case.
The action for the fluctuations is the one in formula (\ref{5daction}).
In the calculations we systematically discard the contributions coming from the UV completion of the field theory, which are power-like terms suppressed as $(r_{h}/r_{s})^n$.

\subsection{Fluctuations}\label{seccalculation}

Following the standard procedure,\cite{Kovtun:2005ev} we assume that the perturbations take a planar wave form in Minkowski space,
$
\delta\Psi(x^\mu,r) = e^{-i(\omega t - q z)} \Psi(r)
$.
The perturbations can be classified according to their transformation under the little group $SO(2)$, which is the remaining symmetry of the system (rotations in the $x-y$ plane). We define
\begin{eqnarray}
\delta g_{tt}(x^\mu,r) &=& e^{-i(\omega t - q z)}  g_{tt}(r) \ch_{tt}(r)\, , \nonumber\\
\delta g_{mn}(x^\mu,r) &=& e^{-i(\omega t - q z)} g_{xx}(r) \ch_{mn}(r)\, , \quad (m,n) \neq (t,t)\nonumber\\
\delta \Phi(x^\mu,r) &=& e^{-i(\omega t - q z)} \cph(r)\, , \nonumber\\
\delta f(x^\mu,r) &=& e^{-i(\omega t - q z)} \cb(r)\,, \nonumber\\
\delta w(x^\mu,r) &=& e^{-i(\omega t - q z)} \cc(r)\, .
\end{eqnarray}
We work in the gauge $\ch_{rm}(r)=0$. The system of perturbations above includes a tensorial mode ($\ch_{xy}$), vectorial modes ($\ch_{tx}$, $\ch_{zx}$, $\ch_{ty}$, $\ch_{zy}$) and scalar modes ($\ch_{tt}$, $\ch_\pperp \equiv \ch_{xx} + \ch_{yy}$, $\ch_{zz}$, $\ch_{tz}$, $\cph$, $\cb$, $\cc$).
Each kind of perturbation can be expressed in term of gauge invariant quantities under the residual gauge symmetry\cite{Kovtun:2005ev,benincasa,mas}
\begin{eqnarray}
\mathrm{Tensorial} & \to & \cz_{T} = \ch_{xy} \, , \nonumber\\
\mathrm{Vectorial} & \to & \cz_{V} = q \ch_{tx} + \omega \ch_{zx} \, , \nonumber\\
\mathrm{Scalar} & \to & \cz_{S} = 2 \ch_{zz} + 4 \frac{q}{\omega} \ch_{tz} - \left[ 1- \frac{q^2}{\omega^2} \frac{g_{tt}'}{g_{xx}'} \right] \ch_\pperp \nonumber \\
&& \qquad\quad + 2 \frac{q^2}{\omega^2} \frac{g_{tt}}{g_{xx}} \ch_{tt} \, ,\nonumber\\
& & \cz_{\cph} = \cph - \frac{\Phi' }{\log'\left[g_{xx}^2 \right]} \ch_\pperp  \, , \nonumber\\
& & \cz_{\cb} = \cb - \frac{f'}{\log'\left[g_{xx}^2 \right]} \ch_\pperp \, , \nonumber\\
& & \cz_{\cc} = \cc - \frac{w'  }{\log'\left[ g_{xx}^2 \right]} \ch_\pperp \, .
\end{eqnarray}
The differential equations for all these fluctuations at the horizon admit solutions behaving as $(\frac{r}{r_h}-1)^{\pm i \frac{\wn T_{0}}{2T}}$, where $\wn = \omega / (2 r_h)$ and $r_h=\pi T_0$ ($T_0$ denotes the temperature of the unflavored ${\cal N}=4$ SYM theory).  The index with negative sign corresponds to incoming wave boundary conditions at the black hole horizon, which in turn give the causal solutions we are interested in.
We impose that the fluctuations vanish at the UV cutoff scale related to $r_s$: the results turn out to be independent of $r_s$ up to suppressed terms in powers of $r_h/r_s$.

\subsection{Tensorial Perturbation}

 We scale $\wn \rightarrow \lambda_{hyd} \wn,\,  \qn \rightarrow \lambda_{hyd} \qn$, where $\qn=q/(2r_h)$ and $\lambda_{hyd}$ is a parameter keeping track of the order of the hydrodynamic expansion. Define
 \be\label{ansatztensorial}
 \cz_T = C_T \left(1-\frac{r_h^4}{r^4}\right)^{-i \frac{\wn T_{0}}{2 T}} \sum_{j=0}^2\sum_{k=0}^2 \cz_T^{j,k}\, \lambda_{hyd}^j\, \epsilon_*^k\,,
 \ee
where higher order terms in $\epsilon_*$ and $\lambda_{hyd}$, which we will not study, are not taken into account.
From the action (\ref{5daction}) we derive the following equation for the perturbation
\begin{equation}\label{eqfortensor}
\cz_{T}''+\frac12 \cz_{T}' \log'{\Bigl(\frac{g_{tt} g_{xx}^3}{g_{rr}}}\Bigr)+ \frac{g_{rr}}{g_{tt}}\Bigl(\omega^2-q^2 \frac{g_{tt}}{g_{xx}}\Bigr)\cz_{T}=0\,.
\end{equation}
With the ansatz (\ref{ansatztensorial}) one can check that the only non-zero term in the solution normalized to one at the horizon up to first order in $\lambda_{hyd}$ is $\cz_{T}^{0,0}=1$.
At second order in $\lambda_{hyd}$ the solution is too lengthy to be reported here but straightforward to obtain.\\

Dirichlet boundary conditions cannot be imposed on the solution in the UV, showing the absence of a hydrodynamic mode in this channel.\cite{Kovtun:2005ev}
We can nevertheless write the hydrodynamic expansion of the retarded correlator once we evaluate the action on-shell.
The latter is singular when evaluated at $r=r_s\to\infty$.
To cure this divergence we have to add the following counterterms\cite{counterterms}
\begin{eqnarray}\label{totalaction}
S_{bulk} &\to & S_{bulk} + \frac{\mathrm{Vol}(X_5)}{2\kappa_{10}^2} \int d^4 \xi \,2\sqrt{-\gamma} \,K \\
&& + \frac{\mathrm{Vol}(X_5)}{2\kappa_{10}^2} \int d^4 \xi \sqrt{-\gamma}\, \Bigl({\cal W[\phi]} - \frac{1}{2} C[\phi]R[\gamma] \Bigr)\,  ,\nonumber
\end{eqnarray}
where $K$ is the scalar associated to the extrinsic curvature of the cut-off hypersurface, $\gamma$ is the four-dimensional metric at the boundary, $C[\phi]$ is a function of the scalars and $\cal W[\phi]$ is the superpotential
\begin{equation}
{\cal W} = e^{\frac{5}{3}f} \left[ Q_c\, e^{5f} + Q_f e^{f-4w+\Phi} - 4 e^{f+6w} - 6 e^{f-4w} \right] \, ,
\end{equation}
from which the potential terms in the action (\ref{5daction}) can be derived as
\begin{equation}
 \frac{1}{2} \left[  \frac{3}{80} \left(\frac{\partial {\cal W} }{\partial f}\right)^2 + \frac{1}{40} \left(\frac{\partial {\cal W} }{\partial w}\right)^2 +  \left(\frac{\partial {\cal W} }{\partial \Phi}\right)^2 \right] - \frac{1}{3} {\cal W} ^2 \, .
\end{equation}
The function $C[\phi]$ satisfies the differential equation\cite{counterterms}
\begin{equation}
\frac{1}{2} - \frac{1}{4} \left[  \frac{3}{80} \frac{\partial {\cal W} }{\partial f}\frac{\partial {C} }{\partial f} + \frac{1}{40}\frac{\partial {\cal W} }{\partial w}\frac{\partial {C} }{\partial w} +  \frac{\partial {\cal W} }{\partial \Phi}\frac{\partial {C} }{\partial \Phi} \right] + \frac{C {\cal W}}{12} = 0 \, .
\end{equation}
The exact form of $C[\phi]$ is actually not needed in order to extract physical results, since the divergence balanced by this function goes in the UV as $r_s^2$, being the next-to-leading order suppressed as $r_s^{-2}$, \emph{i.e.}, it does not affect the finite part from which the hydrodynamic transport coefficients are obtained.
The leading behavior, needed to cancel the divergence, reads
\begin{equation}
C[f,w,\Phi] \approx 1 + \frac{23}{108}\epsilon_* - \frac{371}{23328} \epsilon_*^2 +{\cal O}\left( r^{-4}\right)\,  .
\end{equation}

The Fourier transformed, quadratic-in-fluctuations, on-shell boundary action is
\begin{equation}
S = \frac{\mathrm{Vol}(X_5)}{2\kappa_{10}^2} \int d^4 k H_{-k}{\cal F}(k,r_s) H_k \, ,
\end{equation}
with $H_k$ the boundary value of the fluctuation (${\cal F}$ is obviously not to be confused with the D7-brane world-volume gauge field strength).  The retarded correlator of the corresponding components of the energy momentum tensor
\begin{equation}
G_R^{xy,xy}= - i \int dt d^3x e^{i(\omega t-qz)}\Theta(t) \langle [T_{xy}(t,\vec x), T_{xy}(0,\vec 0)]  \rangle \, ,
\end{equation}
is related to the on-shell action by\cite{Son:2002sd}
\begin{equation}
G_R^{xy,xy} = -2 \,{\rm Im}[{\cal F}(k, r_{s})]\, .
\end{equation}

Plugging the solution of the equation of motion (\ref{eqfortensor}) for $\cz_T$, in the finite action (\ref{totalaction}), it is straightforward to derive the flux ${\cal F}(k,r_s)$,
and so the correlator $G_R^{xy,xy}$ as
\begin{eqnarray}
&& G_R^{xy,xy}= \frac{\pi^5 N_c^2 T^4_{0}}{8 \mathrm{Vol}(X_5)} \Bigl( [1-2 i \wn -2 \qn^2 + 2\wn^2(1-\log{2})] \nonumber\\
&& \qquad \qquad -\frac{ i \wn +2 \qn^2 - 2\wn^2(1-\log{2})}{4}\epsilon_h \\
&& \qquad - \frac{24+19 i \wn -4 \qn^2 + 2\wn^2(2+3\pi^2+22\log{2})}{192}\epsilon_h^2 \Bigr)\,.\nonumber
\end{eqnarray}
Comparing to (\ref{retcorr}) and using the expression for the temperature in (\ref{JT.temperature}) (at zero charge), we confirm the value of the pressure $p$ in section \ref{section:thermodynamics} (at zero charge) and the results in Table \ref{relations} of section \ref{section:coefficients} for the transport coefficients $\eta, \tau_{\pi}, \kappa, \kappa^*$.

\subsection{Vectorial Perturbation}

The equation in this channel reads
\begin{eqnarray}
&& \cz_{V}''+\Bigl[\frac12 \log'{\Bigl(\frac{g_{xx}^5}{g_{tt}g_{rr}}\Bigr)}-\log'{\Bigl(\frac{g_{xx}}{g_{tt}}\Bigr)}\Bigl(1-\frac{q^2}{\omega^2} \frac{g_{tt}}{g_{xx}}\Bigr)^{-1}\Bigr] \cz_{V}' \nonumber \\
&& + \frac{g_{rr}}{g_{tt}}\Bigl(\omega^2-q^2 \frac{g_{tt}}{g_{xx}}\Bigr)\cz_{V}=0\,.
\end{eqnarray}
For the vectorial fluctuation we can solve order by order with the scaling $\wn\to \lambda_{hyd}^2 \wn$,  $\qn\to \lambda_{hyd} \qn$, imposing regularity at the horizon. The result is
\begin{eqnarray}
\cz_V &=& C_V \left(1-  \frac{r_h^4}{r^4}\right)^{-i\frac{\wn T_{0}}{2 T}}  \Bigl[ \frac{r_h^4}{r^4} + \\
&& +\left( 1- i \frac{\qn^2}{\wn} \right)  \left(1-\frac{r_h^4}{r^4} \right) (1+\epsilon_*+\epsilon_*^2) \Bigr] + {\cal O}(\wn,\qn^2) \, .\nonumber
\end{eqnarray}
From Dirichlet conditions at the boundary $r_s\to\infty$ we can read off the shear viscosity from the dispersion relation (\ref{vectorialdispersion}).
This calculation is summarized in the membrane paradigm formula given in Ref. \onlinecite{Kovtun:2003wp}, and it gives the well-known ratio $\eta/s = 1/(4\pi)$ with corrections in powers of $r_h/r_s \to0$.

\subsection{Scalar Perturbations}
We write for each scalar perturbation $\cz_{A=S,\cb,\cc,\varphi}$ the ansatz
\begin{eqnarray}\label{dispersionscalars}
\wn & = & \sum_{k=0}^2 c_{s,k}\,\epsilon_*^k \,\qn- 2\, i\,\sum_{k=0}^2 \gamma_{k}\,\epsilon_*^k \,  \qn^2  + 4\,  \sum_{k=0}^2 t_{k}\,\epsilon_*^k \, \qn^3\,,\nonumber\\
 \cz_A & = & C_A \left(1-\frac{r_h^4}{r^{4}}\right)^{-i\frac{\wn T_0}{2T}} \sum_{j=0}^2\sum_{k=0}^2 \cz_A^{j,k}\, \qn^j\, \epsilon_*^k\,,
\end{eqnarray}
where $\gamma_k, t_k$ are the order $\epsilon_*^k$ coefficients of the dimensionless combinations
\be
\gamma\equiv \pi T_0\, \Gamma\,,\qquad t\equiv (\pi T_0)^2\,\frac{\Gamma}{c_s}\Bigl(c_s^2\tau^{eff}-\frac{\Gamma}{2}\Bigr)\,.
\ee

The relevant equations for the perturbations are
\begin{eqnarray}
&& 2 H_{zz}^{EOM}+4\frac{q}{\omega}H_{tz}^{EOM}-\Bigl(1-\frac{q^2}{\omega^2}\frac{g_{tt}'}{g_{xx}'} \Bigr)H_{aa}^{EOM}\nonumber\\
&&+2\frac{q^2}{\omega^2}\frac{g_{tt}}{g_{xx}}  H_{tt}^{EOM} + \Bigl(\frac{\omega}{q}\frac{g_{xx}}{g_{tt}}\Xi+\frac{4}{\omega}\log'{\frac{g_{xx}}{g_{tt}}} \Bigr)H_{rt}^{EOM}\nonumber\\
&& +\Xi H_{rz}^{EOM}=0\, ,\\
&& \phi^{EOM}-\frac{\phi_B'}{\log'{g_{xx}^2}}H_{aa}^{EOM}\nonumber\\
&&+\frac{\omega g_{xx}[g_{xx}'\phi_B''/2+\phi_B'(\sqrt{g_{xx}}'^2-\sqrt{g_{xx}}\sqrt{g_{xx}}'')]}{\sqrt{g_{xx}}'(q^2 g_{tt}' \sqrt{g_{xx}}/2 +2q^2 g_{tt}\sqrt{g_{xx}}'-3\omega^2 g_{xx}\sqrt{g_{xx}}')} \times \nonumber\\
&& \times \Bigl( H_{rt}^{EOM}+\frac{q g_{tt}}{\omega g_{xx}}H_{rz}^{EOM}\Bigr)=0\, .
\end{eqnarray}
In these expressions, $\phi$ represents each of the scalars $f, w, \Phi$ (the form of their equation is the same) and $\phi_B$ their background value.
Moreover, the notation $\phi^{EOM}$ (and $H_{zz}^{EOM}$ and so on) stands for the corresponding equation for the scalar (and the fluctuation $H_{zz}$ and so on) in section 3 of Ref. \onlinecite{benincasa}.
For the coefficient $\Xi$ we have
\begin{eqnarray}
\Xi &=&-\frac{16q r_h^4(2q^2-3\omega^2)}{r\omega^2[q^2(r_h^4-3r^4)+3r^4\omega^2]}\nonumber\\
&& -\frac{4q^3 r_h^4(q^2-\omega^2)(r^4-r_h^4)}{r\omega^2[q^2(r_h^4-3r^4)+3r^4\omega^2]^2}\epsilon_*^2\, .
\end{eqnarray}

The solutions for the non-zero fluctuations entering the sound channel, satisfying the normalization at the horizon and Dirichlet conditions at the boundary in the case of $\cz_S$, read
 \begin{eqnarray}
 \cz_S^{0,0} & = & \frac{1}{\rho^{4}}\,,\qquad \qquad  \cz_\cph^{0,2}=\frac{\log{\rho}}{12(1-\rho^4)} \,, \\
 \cz_\cph^{1,2} &=& \frac{i}{144\sqrt{3}(1-\rho^4)} \Bigl[ \pi^2(\rho^4-1) +24 (\rho^4-1) \log^2{\rho} \nonumber \\
&& -12 \log{\rho}\Bigl(4+ (\rho^4-1)  \log{(1+i\rho)}+\nonumber \\
 && (\rho^4-1)  \log{[i(i+\rho)(\rho^2-1)]}  \Bigr) -3(\rho^4-1)Li_2(\rho^4)  \Bigl] \nonumber
\end{eqnarray}
where $\rho\equiv r/r_h$. We do not report the expressions for all the $\qn^2$ coefficients of the solutions because of their very lengthy form.\\

The Dirichlet conditions at the boundary give the dispersion relations in the first line of (\ref{dispersionscalars}) with
\begin{eqnarray}\label{res}
c_{s,0} & = & \frac{1}{\sqrt{3}} \, , \qquad \qquad c_{s,1} = 0 \, , \qquad \qquad c_{s,2} = -\frac{1}{12\sqrt{3}} \, , \nonumber \\
\gamma_0 & = & \frac{1}{6} \, , \quad \qquad \gamma_1 = \frac{1}{48} \, , \quad \qquad \gamma_2 = \frac{17-16\log[\frac{r_*}{r_h}]}{768} \, ,\nonumber \\
t_0  & = & \frac{3-2\log{2}}{24\sqrt{3}} \, , \qquad \quad t_1   = \frac{3-2\log{2}}{96\sqrt{3}} \, ,\\
t_2 &=& \frac{57 - 3 \pi^2 - 22 \log{2} - 24 (3 - 2\log{2}) \log[\frac{r_*}{r_h}]}{2304 \sqrt{3}} \, ,\nonumber
 \end{eqnarray}
 which confirm the result for the speed of sound (\ref{speedofsound}) found with the thermodynamics and, using (\ref{vecdiff2}), (\ref{JT.temperature}) and (\ref{JT.gaugerunning}), confirm the results reported in Table \ref{relations} of section \ref{section:coefficients} for the transport coefficients $\zeta, \tau_{\pi}, \tau_{\Pi}$.


\end{document}